\documentclass[reqno,10pt]{amsart}
\usepackage{amssymb,amsthm,color,enumerate,mathrsfs}
\usepackage[colorlinks=true,linkcolor=blue]{hyperref}
\def\fg{\mathfrak{g}}
\def\fl{\mathfrak{L}}

\DeclareMathOperator{\vol}{Vol}
\DeclareMathOperator{\im}{Im}

\numberwithin{equation}{section}
\newtheorem{thm}{Theorem}
\newtheorem{prop}{Proposition}

\theoremstyle{definition}
\newtheorem{rem}{Remark}

\begin{document}
\title{Dilogarithms, OPE and twisted $T$-duality}
\author{Marco Aldi}
\address{Department of Mathematics Brandeis University\\Waltham, MA 02453, USA}
\author{Reimundo Heluani}
\address{IMPA \\ Rio de Janeiro, RJ 22460-320,  Brasil}
\begin{abstract}We study the full sigma model with target the three-dimensional Heisenberg nilmanifold by means of a Hamiltonian formulation of double field theory. We show that the expected $T$-duality with the sigma model on a torus endowed with H-flux is a manifest symmetry of the theory. We compute correlation functions of scalar fields and show that they exhibit dilogarithmic singularities. We show how the reflection and pentagonal identities of the dilogarithm can be interpreted in terms of correlators with 4 and 5 insertions.
\end{abstract}
\maketitle
\section{Introduction} \label{sec:introduction}
This is the first in a series of articles studying the sigma
model with target a nilmanifold, possibly endowed with a gerbe. Our approach to the quantization of the sigma model is intrinsically Hamiltonian and inspired by double field theory, below we will refer to it as Hamitonian Double Field Theory (HDFT).  We focus on the simplest example: the sigma model on a flat three-dimensional torus endowed with $H$-flux or $T$-dually, the sigma model with target the three-dimensional Heisenberg nilmanifold. We prove the equivalence of these models. Finally, we compute HDFT correlators explicitly and show that they can be expressed in terms of Roger's dilogarithms.

Let $\Gamma$ be a rank $2n$ lattice and let $X = \Gamma \backslash \Gamma
\otimes_{\mathbb{Z}} \mathbb{R}$ be the associated torus. Choose a
non-degenerate symmetric
pairing $g=g_{ij}$ on $\Gamma$ and consider the corresponding metric on
$X$, also denoted by $g$. Moreover, let us fix a compatible complex structure on $X$ and $n$ holomorphic coordinates $\alpha^i$ on $X$. The sigma model with target $X$ contains scalar fields
$\alpha^i(z,\bar{z})$, $\bar{\alpha}^i(z,\bar{z})$, subject to the OPE
\begin{equation}
\alpha^i(z,\bar{z}) \cdot \alpha^j(w,\bar{w}) \sim g^{ij} \log|z-w|^2\, ,
\label{eq:1.1}
\end{equation}
where $z$ and $w$ are holomorphic coordinates on a cylindrical worldsheet . While the fields $\alpha^i$ are multivalued, their
derivatives are well-defined and of definite chirality
\begin{equation}
A^i(z):=  \partial_z \alpha^i(z,\bar{z}), \qquad \bar{A}^i(\bar{z}):= \partial_{\bar{z}} \alpha^i(z,\bar{z}).
\label{eq:1.2}
\end{equation}
Differentiation of \eqref{eq:1.1} yields two copies of the Heisenberg vertex algebra $\mathit{Heis}$
\begin{equation}
A^i(z) \cdot A^j(w) \sim \frac{g^{ij}}{(z-w)^2}, \qquad \bar{A}^i(\bar{z}) \cdot \bar{A}^j(\bar{w}) \sim \frac{g^{ij}}{(\bar{z} - \bar{w})^2}.
\label{eq:1.3}
\end{equation}
The phase space of the theory is the cotangent bundle $T^* \mathcal{L}X$ of the free loop space $\mathcal L X$ of the underlying torus. Quantization assigns a Hilbert space of states $\mathcal H$ to the infinite-dimensional symplectic manifold  $T^*\mathcal{L}X$. Geometrically, one can interpret $\mathcal H$ as the completion of a direct sum of Fock modules for the vertex algebra $\mathit{Heis}$  \cite{kapustinorlov}
\def\heis{\mathit{Heis}}
\begin{equation}
\mathcal{H} = \bigoplus_{w \in \Gamma} \mathrm{Ind}_{\heis_+}^{\heis} L^2(X).
\label{eq:1.4}
\end{equation}
Here $w$ parametrizes \emph{winding modes} i.e.\ the connected components of $\mathcal{L}X$. This is how vertex algebras usually arise in the context of quantization, as \emph{the holomorphic part of a two dimensional CFT}.

It is convenient to use a Fourier basis of exponentials to indentify the factor $L^2(X)$ in \eqref{eq:1.4} with a completion of the group algebra $\mathbb{C}[\Gamma^*]$. The vacuum vectors $p \in \Gamma^*$ label quantum states with definite momentum. 

The starting point of our construction is the following observation from double field theory (DFT)\cite{hull}. Since the Lie algebra $\heis$, is spanned by the Fourier modes of $A^i(z)$ and $\bar{A}^i(\bar{z})$ one may obtain a more symmetric expression for \eqref{eq:1.4} by formally introducing operators $W^i$ acting diagonally  (with eigenvalues $w^i$) on quantum states of $\mathcal H$ belonging to a definite winding sector $w$. In other words,
\begin{equation}
\mathcal{H} = \mathrm{Ind}_{\heis_+}^{\heis} \mathbb{C}[\Gamma\oplus\Gamma^*] \,.
\label{eq:1.5}
\end{equation}
The point of view of DFT is that this expression is a consequence of a 1-1 correspondence between the set of \emph{all} vacuum vectors  and the natural Fourier basis of $L^2(Y)$, where $Y:= X\times X^*$ is a $4n$-dimensional torus called the \emph{double torus}. In this picture, the scalar fields $\alpha^i$  and their conjugate momenta $p_i$ correspond to classical coordinates on $T^*X$ as in ordinary canonical quantization. The novelty of DFT is  to interpret the extra coordinates in $Y$ too as honest modes, conjugated to $W^i$.

We now describe our Hamiltonian approach to DFT in the case of tori. In the standard DFT approach, a choice of metric on $X$ is needed in order to write the action and obtain OPEs \eqref{eq:1.1}. Similarly, the expansion of the fields $\alpha^i$ into Fourier modes is a consequence of interpreting those fields as solutions to the equations of motion, which depend on the metric. This should be contrasted with the Hamiltonian approach in which kinematic data, encoded by the symplectic geometry of phase space, are on a fundamentally different footing from the metric content of the theory which describes, through the Hamiltonian operator, the dynamics of the theory. Let us illustrate how HDFT describes the sigma model with target a real torus $X$. For convenience, we choose to realize $X$ as the quotient of $\mathbb R^n$ by the lattice $\Gamma \simeq \mathbb{Z}^n$. The corresponding double torus is then
\begin{equation}\nonumber
Y:=(\Gamma \oplus \Gamma^*)\backslash( (\Gamma\oplus\Gamma^*)\otimes \mathbb R)\cong X\times X^*\,.
\end{equation}
Consider coordinates $(x,x^*)$ on $Y$ obtained by completing a coordinate system $x^i$, $i=1,\dots,n$ on $X$ with coordinates $x^*_i$ on the dual torus $X^*$. Loops in $Y$ can be parametrized as 
\begin{equation}
x^i(\sigma) = 2 \pi \sqrt{-1}\, w^i  \sigma + \sum_{n \in \mathbb{Z}} x^i_n e^{-2 \pi i n \sigma},\quad x^*_i(\sigma) = 2 \pi \sqrt{-1}\, p_i \sigma + \sum_{n \in \mathbb{Z}} x^*_{i,n} e^{-2 \pi i n \sigma}
\label{eq:1.6}
\end{equation}
where $\sigma\in S^1 = \mathbb{Z} \backslash \mathbb{R}$, $(w,p)\in (2 \pi i)^{-1} (\Gamma \oplus \Gamma^*)$, and $(x_0,x^*_0)\in Y$ and the non-zero modes are constrained by requiring the expressions on the LHS to be real. We emphasize that not only  the winding numbers $w^i$, but also the momenta $p_i$ are integral at the classical level. Quantization of this system, requires mapping the coefficients of the expansion \eqref{eq:1.6} to operators acting on a Hilbert space. We begin with a formal change of coordinates $z:=e^{2\pi i \sigma}$ which allows us to rewrite \eqref{eq:1.6} as
\begin{equation}
x^i(z) = w^i \log(z) + \sum_{n \in \mathbb{Z}} x^i_n z^{-n}, \qquad x^*_i(z) = p_i \log(z) + \sum_{n \in \mathbb{Z}} x^*_{i,n} z^{-n}\,.
\label{eq:1.7}
\end{equation}
We introduce operators $W^i$ (resp.\ $P_i$) acting on states of definite winding (resp.\ momentum) by  eigenvalues $w^i$ (resp.\ $p_i$). Furthermore, we impose canonical commutation relations\footnote{To make our formulas more readable, we choose to systematically ignore unitarity issues and omit $\sqrt{-1}$ factors in our commutators.}
\begin{equation}
[P_i,x^j_{0}] = [W^i,x^*_{j,0}]= \delta_i^j, \qquad [x^i_m, x^*_{j,n}] = \frac{1}{n} \delta^i_j \delta_{m,-n} \quad n \neq 0.
\label{eq:1.8}
\end{equation}
The Hilbert space $\mathcal H$ of canonically quantized states is defined as the tensor product of $L^2(Y)$ with the Fock module for the infinite dimensional Heisenberg algebra generated by the non-zero modes. On the first factor, $x^i_0$ and $x^*_{i,0}$ act by multiplication,  $W^i$ acts by differentiation on the $x_i^*$ coordinate and $P_i$ acts by differentiation on the $x^i$ coordinate. We obtain the OPE: 
\begin{equation}
x^*_i(z) \cdot x^j  (w) \sim \delta_i^j \log(z-w).
\label{eq:1.9}
\end{equation}
As a consequence, the Hilbert space $\mathcal{H}$ of HDFT states is naturally identified with (a completion of) the \emph{lattice vertex algebra} associated with $\Gamma \oplus \Gamma^*$. 
The fields \eqref{eq:1.7} are multivalued in the same way as the $\alpha^i$ are. On the other hand, their derivatives are well-defined and can be viewed as extensions of vector fields and differential forms on $X$ to the loop space $\mathcal LX$ or, from the DFT perspective, as extensions of vector fields on $Y$. 

We would like to emphasize the difference between the OPEs \eqref{eq:1.9} and \eqref{eq:1.1}. The point is that \eqref{eq:1.9} contains no dynamical information. Moreover, the coordinate $z$ is  a \emph{formal} parameter and our fields are chiral by construction! This should be contrasted with the standard (if \emph{ante litteram}) DFT approach to lattice vertex algebras of \cite{kapustinorlov} in which the decoupling of holomorphic and anti-holomorphic modes is a property of the solutions to the equations of motions. Furthermore, the HDFT normal ordering determined by the Fourier expansions \eqref{eq:1.7} is different from the one employed in standard DFT, the latter being a consequence of the expansion of $\alpha^i(z,\bar{z})$. Even though both HDFT and standard DFT share the same space of states \eqref{eq:1.5}, the two structures of vertex algebra are a priori unrelated. However, the fact that the Fourier modes of $x^i(z)$ and $x^*_i(z)$ happen to be linear combinations of those of $\alpha^i(z,\bar{z})$ and $\bar \alpha^i(z,\bar z)$ is a hint that, after all, a connection between the two quantization procedures might exist.

A precise dictionary between HDFT and standard DFT requires us to supplement our model with dynamical data i.e.\ to introduce a Hamiltonian operator (cf. \cite{ekstrand} in the vertex algebra context). Let us fix a non-degenerate symmetric pairing $g$ on $\Gamma$ and consider the corresponding metric on $X$, also denoted by $g$. We define an element of the vertex algebra $\mathcal{H}$ by
\begin{equation}
H := \frac{1}{2} g_{ij} :\partial x^i \partial x^j: + \frac{1}{2} g^{ij} :\partial x^*_i \partial x^*_j:
\label{eq:1.10}
\end{equation}
and declare its zero mode $H_0$ to be the Hamiltonian of our theory. Armed with $H$, we can define time evolution for any field $a(\sigma)\in \mathcal H$ by imposing the following equation of motion
\begin{equation}
\frac{\partial}{\partial \tau} a(\sigma,\tau) = [H_0,a(\sigma,\tau)]\, .
\label{eq:1.11}
\end{equation}
Together with \eqref{eq:1.9}, this easily implies
\begin{equation}
\partial_\tau x^i = g^{ij} \partial_\sigma x^*_j, \quad \partial_\tau x^*_i = g_{ij} \partial_\sigma x^j  \Longrightarrow \Bigl( \partial^2_\tau - \partial^2_\sigma \Bigr)x^i(\sigma,\tau) = 0\,.
\label{eq:1.12}
\end{equation}
We deduce that our fields $x_i^*$ and $x^i$ satisfy the same equations of motion as the fields $\alpha^i$. 

We summarize this discussion by restating in HDFT language (the bosonic content of) one of the main results of \cite{kapustinorlov}
\begin{thm}
Let $\Gamma_1, \Gamma_2$ be two lattices. For $i=1,2$, let  $X_i$ be the corresponding tori and let $V_i$ be the associated lattice vertex algebras. The following conditions are equivalent:
\begin{enumerate}[(i)]
\item the sigma models with targets $X_1$ and $X_2$ have isomorphic kinematics (i.e.\ isomorphic canonical quantization),
\item there exists an isomorphisms of vertex algebras $\varphi: V_1 \rightarrow V_2$,
\item there exists and isometry
\begin{equation}\nonumber
\psi: \Bigl( \Gamma_1 \oplus \Gamma_1^*, \langle \cdot,\,\cdot \rangle \Bigr) \rightarrow \Bigl( \Gamma_2 \oplus \Gamma_2^*, \langle \cdot,\,\cdot \rangle \Bigr),
\label{eq:1.13}
\end{equation}
where $\langle,\rangle$ is the tautological symmetric pairing. 
\end{enumerate}
Moreover, given for $i=1,2$ symmetric pairings $g_i$ with associated Hamiltonians $H_i\in V_i$ and isomorphisms $\varphi$ and $\psi$ as above, the following are equivalent: 
\begin{enumerate}[(a)]
\item the sigma models with targets $X_1$ and $X_2$ are isomorphic (i.e.\ they have isomorphic kinematics and equivalent dynamics),
\item $\varphi(H_1) = H_2$,
\item $\psi^*(g_2) = g_1$. 
\end{enumerate}
\label{thm:1.1}
\end{thm}
We illustrate the content of this theorem in the a simplest example. Let $X_1=X_2^*=S^1=\mathbb{Z}\backslash \mathbb{R}$, let  $g_1=R$ be the radius of $X_1$ and let  $g_2=R^{-1}$ be the radius of $X_2$. Then the tautological $T$-duality relating the sigma model on $X_1$ to that on $X_2$ corresponds to the unique isometry $\psi: \mathbb{Z} \leftrightarrow \mathbb{Z}^*$ such that $\psi^*(g_2)=g_1$. 

Let us point out some of the salient features of HDFT that emerge from the study of toroidal sigma models
\begin{enumerate}
\item The basic fields of HDFT are constructed starting from expansions of loops in $Y$ as in \eqref{eq:1.6}. Solving the equations of motion is not necessary at this stage.
\item HDFT fields depend on a single formal parameter $z$ (not on $\bar z$) and give rise to  a natural structure of vertex algebra on the underlying space of states.
\item $T$-duality is a manifest symmetry of HDFT. It is an automorphism of the HDFT vertex algebra that intertwines two Hamiltonian operators.
\end{enumerate}
The interpretation of HDFT correlators requires caution as they do not coincide with the standard ones. This is because standard correlators encode dynamical information that is not accessible to OPEs computed in the HDFT vertex algebra $\mathcal{H}$. In order to recover physical correlators, one needs to solve the equations of motion or the equivalent KZ-type equations imposed on correlators.

It is natural to ask if HDFT is rich enough to encompass the results of \cite{kapustinorlov} on mirror symmetry. We claim that this is indeed the case, though we choose to postpone a detailed analysis to a future publication. The main observation is that the extra datum of a generalized complex structure on $X$ is completely described in purely kinematic terms by promoting the HDFT vertex algebra $\mathcal H$ to a supersymmetric one. While the expression for the SUSY Hamiltonian involves fermions (and the metric), the resulting operator is actually independent on the choice of generalized complex structure.

\subsection{Non-commutative windings}\label{sec:noncommut}
For the bosonic sigma model with target on a torus, we established that HDFT is equivalent to the standard quantization procedure. We now show that the built-in flexibility of HDFT allows us to extend our analysis to a larger class of targets, not readily accessible with standard methods. For sake of concreteness, we consider the first non-trivial example: the sigma-model with target the three-dimensional Heisenberg nilmanifold $X$.  We also consider the sigma model on a three dimensional torus $\tilde X$ with non-trivial $H$-flux, which is believed to be $T$-dual to the sigma model on $X$. Recently, these vacua have been the object of intensive study. We refer the reader to \cite{mathai}, \cite{mathai2} and in particular to \cite{hull2} for a DFT perspective. 

The $3$-dimensional Heisenberg group $H(\mathbb{R})$ is the vector space $\mathbb{R}^3$ with multiplication given by:
\begin{equation}
(x,y,z)\cdot(x',y',z') = \left( x+x', y+y'+\frac{1}{2} x z' - \frac{1}{2} z x',z + z'  \right).
\label{eq:1.14b}
\end{equation}
$H(\mathbb R)$ admits a cocompact lattice $\Gamma$ generated by $e_1$ and $e_3$. The Heisenberg nilmanifold is by definition the quotient $X = \Gamma\backslash H(\mathbb{R})$. We emphasize that the fundamental group $\pi_1(X)\cong\Gamma$ is noncommutative. Connected components of $\mathcal{L}X$ are parametrized by conjugacy classes of $\pi_1(X)$ which implies the following decomposition into topological sectors
\begin{equation}
\mathcal{H}_X \cong \bigoplus_{w \in \pi_1(X)^\sim} \mathcal{H}_{X,w},
\label{eq:1.13b}
\end{equation}
where each $\mathcal{H}_{X,w}$ is a vacuum module generated by $L^2(X)$. 

The sigma model on $X$ contains quantum fields associated with sections of $TX$ and $T^*X$. The triviality of the tangent bundle allows us to construct a global frame of left-invariant sections of $TX\oplus T^*X$
\begin{equation}
\begin{gathered}
X_1 = \partial_{y^1} - \frac{1}{2} y^3 \partial_{y^2}, \quad X_2 = \partial_{y^2},\quad X_3=\partial_{y^3} + \frac{1}{2} y^1 \partial_{y^2},\\
Y^1 = dy^1,\quad  Y^2 =  dy^2 -\frac{1}{2} y^1 dy^3 + \frac{1}{2} y^3 dy^1,\quad Y^3= dy^3
\end{gathered}
\label{eq:1.14}
\end{equation}
where $y^i$ are coordinates on $H(\mathbb R)$. In particular the Lie algebra of vector fields is a copy of the three dimensional Heisenberg algebra with commutators $[X_1,X_3]=X_2$. 

As explained in \cite{alekseev} this Lie algebra, or more generally the Courant-Dorfman algebra of sections $TX\oplus T^*X$, extend to a natural current algebra on the loop space. The explicit trivialization of the Courant algebroid $TX \oplus T^*X$ by $X_i$ and $Y^i$ allows for a concrete description of the corresponding Courant-Dorfman algebra $\fg$ as  the \emph{universal $2$-step nilpotent Lie algebra of rank $3$}.  Given a three dimensional vector space $V$, the associated free $2$-step nilpotent Lie group $G$ is the extension 
\begin{equation}
1 \rightarrow \wedge^2 V \rightarrow G \rightarrow V \rightarrow 1
\label{eq:1.15}
\end{equation}
with group structure
\begin{equation}
(v,\zeta)(v',\zeta') = (v+v', \zeta+\zeta'+v\wedge v'),\qquad v,v' \in V, \quad \zeta,\zeta' \in \wedge^2 V
\label{eq:1.16}
\end{equation}
and Lie algebra $\fg = \mathrm{Lie}(G)$. For later use, we point out that, since $\fg$ has rational structure constants, a theorem of Malcev establishes the existence of a unique (up to isomorphism) cocompact lattice $\Gamma$ in G.
By setting $\beta_1:= X_1$, $\beta_2:=-Y^2$, $\beta_3 := X_3$, $\alpha^1 := Y^1$, $\alpha^2:=-X_2$ and $\alpha^3:=Y^3$ and introducing the totally antisymmetric tensor $\varepsilon_{ijk}$,  the non-trivial commutators of $\fg$ can be written as
\begin{equation}\nonumber
[\beta_i,\beta_j] = \varepsilon_{ijk} \alpha^k\,.
\end{equation}
 Since $\fg$ comes equipped with the left-invariant tautological pairing on $TX \oplus T^*X$, the fields of the theory corresponding to the frame \eqref{eq:1.14} give rise to the affine current vertex algebra associated with $\hat{\fg} = \fg( (t)) \oplus \mathbb{C}$, a.k.a.\ $V^1(\fg)$, with OPEs
\begin{equation}
\beta_i(z) \cdot \beta_j(w) \sim \frac{\varepsilon_{ijk} \alpha^k(w)}{z-w}, \qquad \beta_i(z) \cdot \alpha^j(w) \sim \frac{\delta_{i}^j}{(z-w)^2}.
\label{eq:1.18}
\end{equation}
The Hilbert space of states inherits the structure of $\hat{\fg}$-module.  

In order to apply our HDFT machinery to this model, we first need to find a manifold $Y$ and a natural correspondence between ground states of the theory and elements of a basis of $L^2(Y)$. In \cite{hull2} it it is suggested that $Y$ should be the so-called \emph{double twisted torus}
\begin{equation}\nonumber
Y:=\Gamma \backslash G
\end{equation}
where $\Gamma$ is the cocompact lattice introduce above. In other words, $Y$ is the six-dimensional free $2$-step nilmanifold. In particular, $TY\cong Y\times \fg$.
This motivates us to define the Hilbert space of the theory as
\begin{equation}
\mathcal{H}_X = \mathrm{Ind}_{\fg[ [t]] \oplus \mathbb{C}}^{\hat{\fg}} L^2(Y),
\label{eq:1.19}
\end{equation}
where we induce from the action of $\fg$ on $L^2(Y)$ by derivations. A justification of the identification of $L^2(Y)$ with the ground states of the sigma model on the Heisenberg nilmanifold is offered in appendix \ref{appendix}.

\subsection{The twisted case} \label{sec:twisted} The sigma-model with target $X$ is believed to be $T$-dual to the sigma model with target a three-dimensional torus $\tilde{X}=\mathbb{Z}^3 \backslash \mathbb{R}^3$ endowed with a non-flat $U(1)$-gerbe. The commutativity of $\pi_1(\tilde X)$ simplifies the structure of the topological sectors of the theory and the underlying Hilbert space of states decomposes as
\begin{equation}
\mathcal{H}_{\tilde{X}} \cong \bigoplus_{w \in \mathbb{Z}^3} \mathcal{H}_{\tilde{X},w}\, .
\label{eq:1.20}
\end{equation}
Consider coordinates $\{x^i\}$ on $\tilde{X}$ for which a natural frame of $T\tilde{X} \oplus T^*\tilde{X}$ can be written as
\begin{equation}
\beta_i = \partial_{x^i}, \qquad \alpha^i = dx^i.
\label{eq:1.21}
\end{equation}
It is well known that a $B$-field $B\in \Omega^2(\tilde X)$ deforms the natural Courant-Dorfman bracket  between vector fields $v_1,v_2$ to
\begin{equation}
[v_1,v_2]_{\rm CD} := [v_1,v_2]_{{\rm Lie}} + \iota_{v_1}\iota_{v_2} dB
\end{equation}
where $[-,-]_{{\rm Lie}}$ denotes the usual Lie bracket. Suppose $\tilde X$ is endowed with a $B$-field such that \footnote{With as slight abuse of notation we identify $B$ with $\nu^*B$, where $\nu:\mathbb R^3\to X$ is the natural quotient map. The precise meaning should be clear from the context.}
\begin{equation}
dB=\varepsilon_{ijk} dx^i dx^j dx^k. 
\end{equation}
The span of the frame \eqref{eq:1.21} is closed under the $dB$-twisted Courant-Dorfman bracket, which again identifies with the 2-step nilpotent Lie algebra $\fg$ encountered above. In appendix \ref{appendix} we describe the space of ground states of the twisted sigma model on $\tilde X$ in terms of $2$-holonomy of the gerbe and explain why it is sensible to identify this space with $L^2(Y)$. Roughly speaking, for each winding $w \in \mathbb{Z}^3$ we pick a line bundle $\mathcal{L}_w$ with Chern class given by $i_w dB$. Sections of this bundle comes equipped with a natural $L^2$-norm and we have an isomorphism
\begin{equation}
L^2(Y) \simeq \bigoplus_{w \in \mathbb{Z}^3} L^2 (\mathcal{L}_2).
\label{eq:1.nose}
\end{equation}
Having established a correspondence between the sigma model on $X$ and the gerby sigma model on $\tilde X$ at the level of zero-modes, it is natural to ask:
\begin{enumerate}[(a)]
\item What algebraic structure does HDFT define on the spaces \eqref{eq:1.13b}, \eqref{eq:1.19} and \eqref{eq:1.20} and how does it relate to vertex algebras?
\item Are $T$-dual HDFT algebras naturally isomorphic?
\item If one endows $X$ and $\tilde{X}$ with translation invariant metrics, are the corresponding Hamiltonians intertwined by $T$-duality?
\end{enumerate}
Our work suggests the following answers. In the flat torus case, the scalar fields \eqref{eq:1.9} were obtained by integrating well-defined fields of the current algebra. When a similar integration is carried out here we find that, in addition to \eqref{eq:1.9}, one necessarily has dilogarithmic singularities in the OPEs. This immediately rules out the possibility that our HDFT algebra $\mathcal H$ is a vertex algebra. However, the following algebraic structure is revealed by computing explicit correlation functions. Let $\alpha,\beta,\gamma,\delta\in \mathcal H$. The corresponding four-point functions are 
\begin{equation}
\langle \alpha| \beta(z) \gamma(w) \delta\rangle \in\mathbb{C}( (z))( (w)), \qquad \langle \alpha|\gamma(w) \beta(z) \delta\rangle\in \mathbb{C}( (w))( (z)).
\label{eq:1.22}
\end{equation}
If $\mathcal{H}$ is a vertex algebra\footnote{To be precise, one should $\alpha$ (say) should be viewed as an element of the contragradient module  ${\mathcal H}^{\vee}$.  However, conforming to a widespread abuse of notation, we identify a vertex algebra with the underlying vector space.} then 
\begin{enumerate}
\item both expressions in \eqref{eq:1.22} are expansions in their respective domains of two functions $f$ and $g$ on $\mathbb{C}\mathbb{P}^1\times \mathbb{CP}^1 \setminus \Delta$.
\item $f$  and $g$ are algebraic functions with possibly a pole in the diagonal.\item $f = g$. 
\end{enumerate}
In section \ref{sec:4point} we compute the 4-point functions \eqref{eq:1.22} for certain $\alpha,\beta,\gamma,\delta \in L^2(Y)$. The result shows that an analogue of condition 1) above still holds, with the difference that $f$ and $g$ are now multivalued. More precisely, $f$ and $g$ are  flat sections of a line bundles on $\mathbb{CP}^1$ minus three points ($0$, $\infty$ and $z/w =1$). Moreover, these sections are not algebraic as in 2) but rather analytic (they are exponentials of dilogarithm functions as explained in section \ref{sec:4point}).  Although $f$ and $g$ are sections of different line bundles, there is a natural isomorphism between them which maps $f$ to $g$. Bearing these crucial differences in mind,  the algebraic structure that HDFT defines on $\mathcal H$ is otherwise very much reminiscent of (an analytic version of) the lattice vertex algebra. 
\subsection{Dilogarithms} \label{sec:dilogs} Our correlators exhibit dilogarithmic singularities. We collect here some of the properties of dilogarithms that are relevant to our calculations. Euler's dilogarithm is defined for a real number $0<x<1$ as
\begin{equation}
\mathrm{Li}_2(x) = \sum_{n > 0} \frac{x^n}{n^2}.
\label{eq:1.23}
\end{equation}
A closely related function is Roger's dilogarithm, defined on $(0,1)$ as 
\begin{equation}\nonumber
L(x)= \mathrm{Li}_2(x) + \tfrac{1}{2} \log(x)\log(1-x). 
\end{equation}
Roger's dilogarithm can be extended analytically to $\mathbb{C}\setminus(-\infty,0]\cup [1,\infty)$ using the integral formula
\begin{equation}
L(z) = -\frac{1}{2} \int_0^z \left( \frac{\log(1-x)}{x} + \frac{\log x}{1-x} \right) dx.
\label{eq:1.24}
\end{equation}
Among the several functional identities satisfied by $L$, those relevant to the present discussion are the \emph{reflection identities}
\begin{equation}
L(z) + L (1-z) = L(1) = \frac{\pi^2}{6},  \qquad L(z) + L\left( \frac{1}{z} \right) = 2 L(1),
\label{eq:1.25}
\end{equation}
and the \emph{pentagonal identity}
\begin{equation}
L(x) + L(y) = L(x y) + L \left( \frac{x - xy}{1-xy} \right) + L\left( \frac{y - xy}{1-xy} \right).
\label{eq:1.26}
\end{equation}
Deligne  \cite{deligne} observed that flat sections of certain natural line bundles on $\mathbb{CP}^1 \setminus \{0,1,\infty\}$ can be expressed in terms of dilogarithms. Let $N$ be total space of the unique $\mathbb C^*$-bundle over $\mathbb C^*\times \mathbb C^*$ whose underlying $U(1)$-bundle over $U(1)\times U(1)$ has the three dimensional Heisenberg nilmanifold $X$ as total space. Consider the the projection map
\begin{equation}
\begin{aligned}
\pi: N & \rightarrow \mathbb{C}^* \times \mathbb{C}^* \\
		(x,y,z) &\mapsto \left( e^x, e^z \right).
\end{aligned}
\label{eq:1.28}
\end{equation}
Since $H(\mathbb R)$ is non-commutative, $N$ carries a canonical non-integrable connection $\nabla$. Let $\Sigma$ be a Riemmann surface and and let $f,g$ be two sections of $\mathscr{O}^*_\Sigma$. Pulling back $N$ and $\nabla$ along the map $\Sigma \xrightarrow{(f,g)} \mathbb{C}^*\times\mathbb{C}^*$ we obtain a line bundle $\mathcal{L}_{(f,g)}$ on $\Sigma$ with a connection. Roger's dilogarithm $L(z)$ (or rather it's exponential since we are identifying $\mathbb{C}^* \simeq \mathbb{Z} \backslash \mathbb{C}$)  is then a flat section of this bundle in the particular case when $\Sigma=\mathbb{CP}^1\setminus\{0,1,\infty\}$ and  $g=1-f=z$.  

For general $f,g$, we can trivialize the bundle $\mathcal{L}_{(f,g)}$ by choosing \footnote{Our presentation of the Heisenberg group differs from the ``polarized'' one of Deligne. As a consequence our sections differ by an additive factor of $\tfrac{1}{2} \log(f)\log(g)$ from those of \cite{deligne}. In doing we obtain a trivializing section that does not depend on the choice of $\log(g)$.} logarithms of $f$ and $g$. If we let $s_{f,g}$ be the corresponding trivializing section, then
\begin{equation}
\nabla \cdot s_{f,g} = \log(f) \frac{d g}{g}.
\label{eq:1.29}
\end{equation}
The bundles $\mathcal{L}_{(f,g)}$ are \emph{bimultiplicative} in $f$ and $g$, in the sense that
\begin{equation}\label{bimult}
\mathcal{L}_{(f f',g)} \simeq \mathcal{L}_{(f,g)} \otimes \mathcal{L}_{(f',g)}\qquad \mathcal{L}_{(f,g g')} \simeq \mathcal{L}_{(f,g)} \otimes \mathcal{L}_{(f,g')}\,.
\end{equation}
 As a consequence, one obtains the trivializing section
\begin{equation}
s_{f,g} + s_{g,f} \in \Gamma \Bigl( \mathcal{L}_{(f,g)} \otimes \mathcal{L}_{(g,f)} \Bigr)
\label{eq:1.30}
\end{equation}
which in independent on the choice of $\log(f)$ and $\log(g)$. Similarly,  $s_{f,f} \mapsto s_{f,-1}$ defines natural isomorphisms
\begin{equation}
\mathcal{L}_{(f,f)} \simeq \mathcal{L}_{(f,-1)} \simeq \mathcal{L}_{(-1,f)}^{-1} \qquad \mathcal{L}_{(-f,f)} \simeq \mathscr{O}
\label{eq:1.30b}
\end{equation}
which are well-defined independently on the choice of $\log(f)$. Combining \eqref{bimult}, \eqref{eq:1.30} and \eqref{eq:1.30b} we obtain
\begin{multline}
\mathcal{L}_{(1-f^{-1},f^{-1})}^{-1} \simeq \mathcal{L}_{(1-f^{-1},f)} \simeq \mathcal{L}_{(-f^{-1},f)} \otimes \mathcal{L}_{(1-f,f)} \simeq \mathcal{L}_{(-f,f)} \otimes \mathcal{L}_{(1-f,f)} \simeq \mathcal{L}_{(1-f,f)}
\label{eq:1.30c}
\end{multline}

In $f+g = 1$, Roger's dilogarithm produces trivializations $s_{1-g,g}$ of $\mathcal{L}_{(f,g)} = \mathcal{L}_{(1-g,g)}$ and $s_{1-f,f}$ of $\mathcal{L}_{(g,f)} = \mathcal{L}_{(1-f,f)}$. It follows from the first reflection identity \eqref{eq:1.25} that 
\begin{equation}\nonumber
s_{1-g,g}+s_{1-f,f}+L(1)
\end{equation}
is a trivialization of $\mathcal{L}_{(f,g)} \otimes \mathcal{L}_{(g,f)}$. Similarly, fixing a trivialization $s_{1-f^{-1},f^{-1}}$ of $\mathcal{L}_{(1-f^{-1},f^{-1})}$, the second reflection identity in \eqref{eq:1.25} implies that 
\begin{equation}\nonumber
s_{1-f^{-1}.f^{-1}}+s_{1-f,f}-2L(1)
\end{equation}
is the trivialization  corresponding to the isomorphism between the first and the last term of $\eqref{eq:1.30c}$. 

Furthermore, as a consequence of \eqref{bimult} we obtain
\begin{equation}
\mathcal{L}_{(1-f,f)} \otimes \mathcal{L}_{(1-g,g)} \simeq \mathcal{L}_{(1-fg,fg)} \otimes \mathcal{L}_{\left( 1-\frac{f-fg}{1-fg},\frac{f-fg}{1-fg} \right)} \otimes \mathcal{L}_{\left( 1-\frac{g-fg}{1-fg},\frac{g-fg}{1-fg} \right)} \,.
\label{eq:1.31}
\end{equation}
The associated canonical trivialization is provided by the pentagonal identity \eqref{eq:1.26}.

For the purpose of this paper, one more geometric characterization of dilogarithms needs to be addressed. Let $\mathbb{H}_3 = \mathbb{C} \times \mathbb{R}_+$ be Lobachevsky's space with its standard Hyperbolic metric, so that geodesics are either vertical lines or semicircles in vertical planes and endpoints in $\mathbb{C} \times \{0\}$. An \emph{ideal tetrahedron} is a geodesic tetrahedron $\Delta$ in $\mathbb{H}_3$ with vertices on the boundary 
\begin{equation}\noindent
\partial \mathbb{H}_3=\mathbb{C} \cup \{\infty\} = \mathbb{CP}^1\,.  
\end{equation}
As explained in \cite{zagier}, if $z_i \in \mathbb{CP}^1$ $i=1,\dots,4$ denote the location of the vertices of $\Delta$, the hyperbolic volume of $\Delta$ is computed as the imaginary part of the dilogarithm\footnote{note that the real part is an elementary function.} 
\begin{equation}
\vol (\Delta) = \im L\left( \frac{(z_1 - z_3)(z_2 - z_4)}{(z_1 -z_4) (z_2 - z_3)} \right).
\label{eq:1.16b}
\end{equation}
Since any compact oriented hyperbolic $3$-manifold with finite volume can be triangulated (possibly after removing a finite number of geodesics) by finitely many hyperbolic tetrahedra, it follows that the volume of any such manifold is a sum of dilogarithms. As we shall see, the basic HDFT correlators are naturally described in terms of volumes of ideal tetrahedra. In the following we abuse notation by omitting the imaginary part from the RHS of \eqref{eq:1.16b}.

\subsection{HDFT Correlators as sections of line bundles} As we pointed out, the space of HDFT states \eqref{eq:1.19} cannot be a vertex algebra because the HDFT correlation functions are not algebraic functions. As explained in section \ref{sec:4point}, HDFT 4-point functions  corresponding to exponentials $\alpha,\beta,\gamma,\delta \in L^2(Y)$ are line bundles $\mathcal{L}_{(f,g)}$ over $\mathbb{CP}^1\setminus\{0,1,\infty\}$ described in the previous section. The understanding of the full structure of $\mathcal H$ requires the calculation of more general correlators involving fields associated with a suitable smooth basis of $L^2(Y)$. One natural choice of such a basis is in terms of theta-functions along the lines described in appendix \ref{appendix}. A more detailed account of the structure of $\mathcal H$ will be given elsewhere. Our main focus here is to observe that dilogarithmic singularities already arise in the calculation of corrrelators associated with the standard Fourier basis of $L^2(Y)$. More precisely, let $x^*_{i}$ $i=1,2,3$ be the coordinates along the fibers of the torus fibration underlying $Y$ and let  $\alpha_w = \exp(\sum_i w^i x^*_i)$, where the tuple $\{w^i\} \in 2 \pi \sqrt{-1} \cdot \mathbb{Z}^3$. Moreover, consider vectors $w_j \in 2 \pi \sqrt{-1} \cdot \mathbb{Z}^3$ $j=1,2,3$, the corresponding functions $\alpha_{i}=\alpha_{w_i}$ and their product $\beta=\alpha_1 \alpha_2 \alpha_3$. As shown in section \ref{sec:4point} 
\begin{equation}
\langle \beta | \alpha_1 (z) \alpha_2(w) \alpha_3 \rangle = \exp\left( \mathrm{vol}(w_1,w_2,w_3) L\left( \frac{w}{z} \right) \right),
\label{eq:1.32}
\end{equation}
where the RHS is defined in the domain $|z|\gg |w|$ and $\mathrm{vol}$ denotes the (oriented) volume of the Euclidean tetrahedron spanned by the three vectors $w_i$. This realizes the correlator \eqref{eq:1.32} as a flat section (with argument $w/z\notin\{0,1,\infty\}$ of the line bundle $\mathcal{L}_{(1-f,f)}$ over $\mathbb{CP}^1\setminus\{0,1,\infty\}$, where $f=\frac{w}{z}$.  Permuting $\alpha_{w_1}(z) \leftrightarrow \alpha_{w_2}(w)$ on the LHS of \eqref{eq:1.32} is reflected on the RHS by changing the sign of the Euclidean volume by replacing the argument of the dilogarithm with its reciprocal $z/w$ . Therefore, we see that $\langle \beta| \alpha_2(w)\alpha_1(z)\alpha_3\rangle$ is the expansion in the domain $|w|\gg |z|$ of a flat section of the bundle $\mathcal{L}^{-1}_{(1-f^{-1},f^{-1})}$ which is identified with \eqref{eq:1.32} by the isomorphism \eqref{eq:1.30c}. 

As anticipated, this justifies the following reformulation of conditions 1)-3) of section \ref{sec:twisted}. 
\begin{enumerate}
\item The four point functions \eqref{eq:1.22} converge in their respective domains to flat sections of line bundles on $\mathbb{CP}^1 \setminus \{0,1,\infty\}$.
\item These bundles are naturally isomorphic and under these isomorphisms the sections given by the two $4$-point functions are identified. 
\end{enumerate}

\textbf{Acknowledgements:} The authors are indebted to Edward Frenkel and Maxim Zabzine for illuminating discussions. The work of M.A.\ is supported by NSF FRG grant DMS-0854965.
\section{$3$-point functions, definition of the fields}
The \emph{double twisted torus} $Y$ is defined as the quotient of the $6$-dimensional unipotent group \eqref{eq:1.15} modulo the subgroup $\Gamma$ generated by $e_i$, where $\{e_1,e_2,e_3\}$ is the standard basis of $V = \mathbb{R}^3$. We introduce coordinates $\{x^i, x^*_i\}$, $i=1,2,3$ on $G$ such that the group law is given by
\begin{equation}
(x^i,x^*_i) \cdot (y^i,y^*_i) = \left(x^i+y^i, x^*_i + y^*_i + \frac{1}{2} \varepsilon_{ijk} x^j y^k \right),
\label{eq:2.1}
\end{equation}
where $\varepsilon$ is the totally antisymmetric tensor. The tangent bundle $TY$ is trivialized by the left invariant vector fields of $G$, which in this coordinate system are:
\begin{equation}
\alpha^i = \frac{\partial}{\partial x^*_i}, \qquad \beta_i = \frac{\partial}{\partial x^i} - \frac{1}{2} \varepsilon_{ijk} x^j \frac{\partial}{\partial x^*_k}, \qquad i=1,2,3.
\label{eq:2.2}
\end{equation}
The vector fields $\alpha^i$ are well defined because $Y$ is a $T^3$ fibration over $T^3$ with coordinates $x^*_i$ along the fibers and coordinates $x^i$ on the base. The six vectors $\beta_i$, $\alpha^i$ with commutators 
\begin{equation}
[\beta_i, \beta_j] = \varepsilon_{ijk} \alpha^k.
\label{eq:2.3}
\end{equation}
span the Lie algebra $\fg$. Moreover, $\fg$ carries a non-degenerate invariant symmetric form\begin{equation}
(\beta_i, \alpha^j) = \delta_i^j.
\label{eq:2.4}
\end{equation}
This metric induces an isomorphism $\pi:TY \overset{\sim}{\rightarrow} T^*Y$ such that\begin{equation}
\pi (\alpha^i) = dx^i, \qquad \pi(\beta_i) := dx^*_i -\frac{1}{2} \varepsilon_{ijk} x^j dx^k. 
\label{eq:2.5}
\end{equation}
Each current of the associated vertex algebra $V^1(g)$ of \eqref{eq:1.18}, corresponds to a vector field on $Y$ and their OPEs extend the Lie bracket. Alternatively, using $\pi$, we may think these currents as associated with differential forms on $Y$. We now proceed to \emph{integrate} these fields to obtain scalar fields associated with the coordinates $\{x^i,x^*_i\}$ as done for the double torus. Ideally, one would like to obtain possibly multivalued fields $x^i(z)$ and $x^*_i(z)$ whose worldsheet derivatives are the well-defined currents $\alpha^i$ and $\beta_i$.  For the coordinates along the base this can be done because
\begin{equation}
\partial_z x^i(z) = \alpha^i(z).
\label{eq:2.6}
\end{equation}
This is not surprising since  $x^i(z)$ is the field associated with the coordinate $x^i$ on $Y$, and its derivative should be associated with the differential form, $dx^i$ In turn, using the isomorphism $\pi$ this is further identified with the current $\alpha^i$. 

Unfortunately, integration is less straightforward for $x^*_i$ as the corresponding differential form$dx^*_i$ is not well defined. A quick look at \eqref{eq:2.5} or \eqref{eq:2.2} shows that it is natural to impose then
\begin{equation}
\partial_z x^*_i(z) = \beta_i(z) + \frac{1}{2} \varepsilon_{ijk} :x^j(z) \partial_z x^k(z):\,.
\label{eq:2.7}
\end{equation}
Actually, we may avoid the normal ordering since the modes of $x^i(z)$ commute. In what follows, we describe these multivalued fields and their HDFT OPEs in more detail. Logarithms of $z$ are unavoidable in the Fourier expansions of the HDFT fields and they appear already in the torus case, as \eqref{eq:1.7} shows. Perhaps surprisingly, $\log$ is the only multivalued function needed in this twisted case as well. Our goal is to find explicit ${\rm End}(\mathcal{H})$-valued expansions for the fields $x^i(z)$ and $x^*_i(z)$, impose canonical commutation relations  and obtain OPEs among those fields. Before embarking in the details of this calculation, we first give a quick but less rigorous way of obtaining these OPE. 
\subsection{Heuristics: formal integration} Somewhat naively, we would like to compute the OPE of the fields $x^i(z)$ and $x^*_i(z)$ by integrating \eqref{eq:1.18} and using \eqref{eq:2.6} together with \eqref{eq:2.7}.
Since the fields $\alpha^i$ commute, 
\begin{equation}
x^i(z) \cdot x^j(w) \sim 0\, .
\label{eq:2.8}
\end{equation}
In analogy with the treatment given for tori, integrating the second equation of \eqref{eq:1.18} yields \eqref{eq:1.9}. This implies that we may use the same expansion for $x^i(z)$ as in \eqref{eq:1.7}. Even though the notation of \eqref{eq:1.9} is quite common in the physics literature, later we would like to work in the setting of formal distributions where we may allow now logarithms in the Fourier coefficients. Whenever we write a function $f(z,w)$ in the singular part of the OPE we will mean to take the expansion $i_{z,w}f(z,w)$ in the domain $|z|\gg |w|$ and then subtract the expansion $i_{w,z} f(w,z)$. We note that this is slightly trickier than what is usually done in vertex algebra theory where the only singularities allowed for $f$ are of the form $(z-w)^{-n}$. Explicitly, by \eqref{eq:1.9} we mean
\begin{multline}
[x^*_i(z), x^j(w)] = \delta_{i}^j i_{z,w} \log(z-w) -\delta_i^j  i_{w,z} \log(w-z) =\\ \delta_i^j \left( \log(z) + i_{z,w} \log\left( 1 - \frac{w}{z}  \right) \right) - \delta_i^j \left( \log(w) + i_{w,z} \log\left( 1-\frac{z}{w} \right)\right) = \\  \delta_i^j \left( \log(z) - \sum_{n > 0} \frac{w^n z^{-n}}{n}\right) - \delta_i^j \left( \log(w) - \sum_{n > 0} \frac{z^{n} w^{-n}}{n}\right) = \\ \delta_i^j \left( \log\left( \frac{z}{w} \right) + \sum_{n \neq 0} \frac{z^n w^{-n}}{n} \right), 
\label{eq:2.8b}
\end{multline}
as an equality of formal distributions. What distinguishes our HDFT algebra $\mathcal H$ from a lattice vertex algebra is the OPE of the fields $x^*_i$. From the first equation of \eqref{eq:1.18} and \eqref{eq:2.7} we obtain
\begin{equation}
\partial_zx^*_i(z) \cdot \partial_w x^*_j(w) \sim - \frac{1}{2} \varepsilon_{ijk} \frac{x^k(z) - x^k(w)}{(z-w)^2}. 
\label{eq:2.9}
\end{equation}
Performing the first integration
\begin{equation}
x^*_i(z)\cdot x^*_j(w) \sim - \frac{1}{2} \varepsilon_{ijk} \int \frac{x^k(w) dw + x^k(z) dz}{z-w}.
\label{eq:2.10}
\end{equation}
When integrating \eqref{eq:2.10} it might be tempting to use 
\begin{equation}
\frac{x^k(z) - x^k(w)}{z-w} = (x^k(z) - x^k(w)) \delta(z-w) = 0\,.
\label{eq:2.11}
\end{equation}
However, since we would like to work with formal distributions care must  be taken. The expansion of $x^k$ contains logarithms \eqref{eq:2.11}  and would imply the equation $\delta(z-w) \log(z/w) = 0$. To solve this problem we separate the multivalued part of the field $x^i$ defining
\begin{equation}
\tilde{x}^i(z) = x^i(z) - W^i \log(z),
\label{eq:2.12}
\end{equation}
to obtain
\begin{multline}
x^*_i(z) \cdot x^*_j(w) \sim - \frac{1}{2} \varepsilon_{ijk} \int \frac{\tilde{x}^k(z) dw + \tilde{x}^k(w) dz}{z-w}  - \\  \frac{1}{2} \varepsilon_{ijk} W^k \int \frac{\log(w) dw + \log(z) dz}{z-w} = \\  \frac{1}{2} \varepsilon_{ijk} \log(z-w) \bigl( \tilde{x}^k(z) - \tilde{x}^k(w) \bigr) +  \varepsilon_{ijk} L\left( \frac{w}{z} \right) W^k + 2 \varepsilon_{ijk} L(1) W^k
\label{eq:2.13}
\end{multline}
As above, this equation is the commutator of the fields is the difference of the expansions of $\log$ and Roger's dilogarithm $L$ in the domains $|z| \gg |w|$ and $|w| \gg |z|$. Of course, since we are just formally integrating \eqref{eq:1.18} and the only equation we are imposing is \eqref{eq:2.7}, the constants of integration are arbitrary. In particular, the last term involving $L(1)$ in \eqref{eq:2.13} is there for compatibility with the computations below. In fact we will see that the constant term on the commutator $[x^*_i(z),x^*_j(w)]$ is zero therefore the appearance of $L(1)$ here is due to \eqref{eq:1.25}. 

The importance of \eqref{eq:2.13} is that in the twisted situation we are \emph{forced} to have dilogarithmic singularities in the OPE of scalar fields. From the point of view of the three torus with coordinates $x^i$ endowed with an $H$-flux given by $\varepsilon_{ijk}$, the operators $W^i$ are winding operators, so if we restrict to the winding zero sector of the theory we will not see these singularities.  If we apply $T$-duality to the second coordinate, we obtain the sigma model with target the Heisenberg nilmanifold,  the operator $W^2$ now measures momentum in the central direction (the second coordinate is along the center of the Heisenberg group \eqref{eq:1.14b} in our picture) so in this picture, even restricting to trivial windings, we are forced to consider dilogarithmic singularities. 

We conclude by writing an alternative formulation of \eqref{eq:2.13} . Using  
\begin{equation}
\partial_w  \log(z-w) \tilde{x}^k(w)  = - \frac{\tilde{x}^k(w)}{z-w} +  \log(z-w) \partial_w \tilde{x}^k(w),
\label{eq:2.13b}
\end{equation}
we obtain the convenient expression:
\begin{equation}
x^*_{i}(z) \cdot x^*_j(w) \sim \varepsilon_{ijk} \left[ W^k L\left( \frac{w}{z}  \right) + 2 W^k L(1) - \frac{1}{2} \int \frac{\partial_w \tilde{x}^k(w)}{z-w} dz dw\right].
\label{eq:2.13c}
\end{equation}
\subsection{Loops on $Y$}
We now would like to show that \eqref{eq:2.13} can be rigorously deduced in HDFT. One of the advantages of HDFT is we do not need to solve the equations of motion but simply find mode expansions for loops on the double twisted torus $Y$. Choosing a coordinate $\sigma$ on $S^1 = \mathbb{Z}\backslash \mathbb{R}$, elements of $\mathcal Y$ can be written as
\begin{equation}
\begin{aligned}
x^i(\sigma) &= 2 \pi \sqrt{-1} W^i \sigma + \sum_{n \in \mathbb{Z}} x^i_n e^{-2 \pi i n \sigma} \\ 
x^*_i(\sigma) &= 2 \pi \sqrt{-1} P_i \sigma + \sum_{n \in \mathbb{Z}} x^*_{i,n} e^{-2n\pi i \sigma} + \frac{2 \pi \sqrt{-1} \sigma}{2} \varepsilon_{ijk} W^j x^k(\sigma)
\end{aligned}
\label{eq:2.15}
\end{equation}
where the tuple $(W^i, P_i)$ is an element of $(2 \pi \sqrt{-1})^{-1} \Gamma$, the tuple of real numbers $(x^i_0, x^*_{i,0})$ is well defined modulo $\Gamma$ and the other coefficients are constrained solely by the requirement that the expressions on the LHS are real numbers.   
As $\sigma \mapsto \sigma+1$ that the tuple $\bigl(x^i(\sigma+1), x^*_i(\sigma +1)\bigr)$ is given by
\begin{multline}
(x^i(\sigma) + 2 \pi \sqrt{-1} W^i, x^*_i(\sigma) + 2 \pi \sqrt{-1} P_i + \frac{1}{2} 2 \pi \sqrt{-1} \varepsilon_{ijk} W^j x^k(\sigma)) = \\ \bigl( 2 \pi \sqrt{-1} W^i, 2 \pi \sqrt{-1} P_i \bigr) \cdot \bigl( x^i(\sigma), x^*_i(\sigma) \bigr).
\label{eq:2.16}
\end{multline}
so our fields are well defined. Making the change of coordinates $z= e^{2 \pi i \sigma}$ results in the expansions:
\begin{equation}
\begin{aligned}
x^i(z) &= W^i \log(z) + \sum_{n \in \mathbb{Z}} x^i_n z^{-n}, \\ x^*_{i}(z) &= P_i \log(z) + \sum_{n \in \mathbb{Z}} x^*_{i,n} z^{-n} + \frac{\log(z)}{2} \varepsilon_{ijk} W^j x^k(z)
\label{eq:2.17}
\end{aligned}
\end{equation}
Upon quantization, we would like to replace each mode of these fields by operators acting on $\mathcal{H}$. The basic HDFT fields $x^i(z)$, $x_i^*(z)$ are multivalued but their derivatives \eqref{eq:2.6} and \eqref{eq:2.7}  are well-defined and generate the current algebra \eqref{eq:1.18}. From \eqref{eq:2.8} it is natural to impose that all the coefficients $x^i_n$ commute with themselves and with $W^i$ for all $i$ and $n$. From \ref{eq:2.17} we see that the second equation of \eqref{eq:1.18} is equivalent to 
\begin{equation}
[\partial_z x^*_{i}(z), \partial_w x^j(w)] = \delta_i^j \partial_w \delta(z-w).
\label{eq:2.19}
\end{equation}
Comparing the coefficients of $z^{-n-1} w^{-1-m}$ on both sides of \eqref{eq:2.19} we obtain the following commutation relations\footnote{We include $\hbar$ here to obtain a Lie algebra. Later in the text we will put $\hbar=1$. To work in a unitary setting one should replace $\hbar \mapsto \sqrt{-1} \hbar$ in these commutations.}:
\begin{equation}
\begin{aligned}
{[}P_i,W^j] &= [P_i, x^j_{m}] = [W^j,x^*_{i,m}] = 0, && m \neq 0 \\
[x^*_{i,n}, x^j_m] &= \frac{1}{m} \delta_i^j \delta_{m,-n} \hbar, &&  m \neq 0.
\end{aligned}
\label{eq:2.20}
\end{equation}
Imposing the first equation of \eqref{eq:1.18} we obtain 
\begin{equation}
\begin{aligned}
{[}P_i, P_j] &= - \varepsilon_{ijk} W^k, \\
[P_i, x^*_{j,n}] &= - \frac{1}{2} \varepsilon_{ijk} x^k_n, && n \neq 0, \\
[x^*_{i,m}, x^*_{j,n}] &= \frac{m + n}{2 mn} \varepsilon_{ijk} x^k_{m+n} + \frac{1}{m^2} \varepsilon_{ijk} W^k \delta_{m,-n}, && m,n \neq 0
\end{aligned}
\label{eq:2.21}
\end{equation}
We recognize the last term of the last commutator as the quadratic factor responsible for the appearance of the dilogarithm in \eqref{eq:2.13}. Also we notice that the commutator of $P_i$ and $W^j$ as \emph{right} invariant vectors of $G$  vanishes for all $i$ and $j$ (changing the sign of $\varepsilon$ in \eqref{eq:2.2}). Therefore, may think of these operators as generators of the infinitesimal action of $G$ on $Y = \Gamma \backslash G$ (on the right) thus on functions on $Y$.  

So far we only imposed constraints on the derivatives of the fields $x^i(z)$ and $x^*_i(z)$. We now need to impose commutation relations for the zero modes of our model. According to the DFT philosophy, the operators $x^*_{i,0}$ and $x^i_0$ are naturally associated with coordinate functions on $Y$ which makes it natural to quantize them by prescribing canonical commutation relations with the corresponding conjugate momenta:
\begin{equation}
\begin{aligned}
{[}P_i, x^j_0] &= [W^j, x^*_{i,0}] = \delta_i^j \hbar. \\
[P_i, x^*_{j,0}] &= - \frac{1}{2}\varepsilon_{ijk} x^k_0.
\label{eq:2.18}
\end{aligned}
\end{equation}
In addition impose the $n=0$ limiting case of \eqref{eq:2.21}
\begin{equation}
[x^*_{i,m}, x^*_{j,0}] = \frac{1}{2 m} \varepsilon_{ijk} x^k_m, \qquad m \neq 0.
\label{eq:2.22}
\end{equation}
This exhausts the list of the non-vanishing commutation relations we impose. In other words, our interested is in representations $\mathcal{H}$ of the Lie algebra $\fl$ spanned by $\hbar, W^i$, $P_i$, $x^i_n$, $x^*_{i,n}$, $i=1,2,3$, $n \in \mathbb{Z}$ with relations \eqref{eq:2.20}, \eqref{eq:2.21}, \eqref{eq:2.18} and \eqref{eq:2.22} (and all other brackets between generators set to zero). To check that  $\fl$ satisfies the Jacobi identity, consider first the case $m,n,o \neq 0$
\begin{equation}
[x^*_{k,o}, [x^*_{i,m},x^*_{j,n}]] = \frac{1}{2 m n} \delta_{m + n + o, 0} \hbar,
\label{eq:2.23}
\end{equation}
so that the Jacobiator is 
\begin{equation}
\frac{\varepsilon_{ijk}}{2} \left( \frac{1}{mn} + \frac{1}{no} + \frac{1}{om} \right) \delta_{m+n+o,0} \hbar= 0
\label{eq:2.24}
\end{equation}
When $m,n \neq 0$ 
\begin{equation}
[x^*_{k,0}, [x^*_{i,m}, x^*_{j,n}]] = - \frac{1}{m^2} \varepsilon_{ijk} \delta_{m,-n}\hbar, \quad [x^*_{i,m}, [x^*_{j,n}, x^*_{k,0}]] = \frac{1}{2 m^2} \varepsilon_{ijk} \delta_{m,-n} \hbar
\label{eq:2.25}
\end{equation}
and again the Jacobiator vanishes. Finally, if $m = n = 0$ then all triple brackets are zero.  The checks involving $P_i$ and $x^*_{j,m}$ are left to the reader. 

The Lie algebra $\fl$ is $\mathbb{Z}$-graded where the degree $0$ part is the $13$-dimensional Lie algebra spanned by $\hbar$, $P_i$, $W^i$ and their conjugate coordinates $x^i_0$ and $x^*_{i,0}$. The remaining generators $x^i_n$ and $x^*_{i,n}$ are of degree $n$. In particular, we have a triangular decomposition 
\begin{equation}
\fl = \fl_- \oplus \fl_0 \oplus \fl_+. 
\label{eq:2.26}
\end{equation}
As noticed above, it makes sense to consider the action of $G$ on $L^2(Y)$ by right translations whose differential gives us an action of $\fl_0$ on $C^\infty(Y)$ with $\hbar$ acts as $1$, $x^*_{i,0}$ and $x^i_0$ acting by multiplication and $P_i$, $W^i$ acting by the derivations
\begin{equation}
P_i = \partial_{x^i} + \frac{1}{2} \varepsilon_{ijk} x^j \partial_{x^*_k}, \qquad W^i = \partial_{x^*_i}.
\label{eq:2.27}
\end{equation}
We extend this representation to all of $\fl_+$ by requiring the positive modes to anihilate $C^\infty(Y)$, define
\begin{equation}
\mathcal{H} := \mathrm{Ind}^{\fl}_{\fl_+} C^\infty(Y), 
\label{eq:2.28}
\end{equation}
and complete this to a Hilbert space. This notation is justified as $\mathcal H$  is naturally isomorphic to the space of states introduced  in \eqref{eq:1.19}. To see this, we express all the generators of $\fl$ in terms of the generators of $\hat{\fg}$, except for the zero modes which we add by hand. Indeed, consider the vertex algebra $V^1(\fg)$ generated by fields
\begin{equation}
\beta_i(z) = \sum \beta_{i,n} z^{-1-n}, \qquad \alpha^i(z) = \sum \alpha^i_n z^{-1-n},
\label{eq:2.nose1}
\end{equation}
with OPE given by \eqref{eq:1.18}, which is equivalent to
\begin{equation}
[\beta_{i,m}, \beta_{j,n}] = \varepsilon_{ijk} \alpha^k_{m+n}, \qquad [\beta_{i,m}, \alpha^j_{n}] = m \delta_{i}^j \delta_{n,-m}
\label{eq:2.nose2}
\end{equation}
From the first equation in \eqref{eq:2.17} and \eqref{eq:2.6} we obtain
\begin{equation}
x^i_n = - \frac{1}{n} \alpha_n, \qquad W^i = \alpha_0. 
\end{equation}
We add the zero mode $x^j_0$ by imposing
\begin{equation}
[\beta_{i,n}, x^j_0] = \delta_{i}^j \delta_{n,0}.
\label{eq:2.nose3}
\end{equation}
A direct computation using \eqref{eq:2.17} and \eqref{eq:2.7} shows
\begin{equation}
\begin{aligned}
x^*_{i,n} &= -  \frac{1}{n} \left( \beta_{i,n} + \varepsilon_{ijk} x^j_n W^k - \frac{1}{2} \varepsilon_{ijk} \sum_{m \in \mathbb{Z}} m x^j_{n-m} x^k_m  \right) \\
P_i &= \beta_{i,0} + \varepsilon_{ijk} x^j_0 W^k - \frac{1}{2} \varepsilon_{ijk} \sum_{m \in \mathbb{Z}} m x^j_{-m} x^k_m.
\end{aligned}
\label{eq:2nose4}
\end{equation}
It is natural to ask if there is a vertex algebra associated with this infinite dimensional Lie algebra. The natural candidate would be its vacuum representation, but this coincides with the vacuum representation of $\hat{\fg}$. What we learn is that the formalism of vertex algebras fails here essentially because the basic HDFT fields  \eqref{eq:2.17} contain logarithms, preventing $\fl$ from being a formal distribution Lie algebra. 

Notice that in toroidal models too one has logarithms in the expansion of the basic fields  toroidal models. However, these models are simpler because only one logarithm  is present in each expansion (say of $x^*_{i}(z)$) which offers a natural way out i.e.\ to work with exponentiated fields\begin{equation}
:\exp( x^*_{i}(z)): 
\label{eq:2.29}
\end{equation}
where $:\,:$ denotes normal ordering. This is in stark contrast with the expansions  \eqref{eq:2.17} which involve infinitely many logarithmic terms in the expansion of $x^*_i(z)$. 

We are now in position to show how \eqref{eq:2.13} and \eqref{eq:2.13c} are a consequence of our expansions. Let us define the anihilation and creation parts of the fields as
\begin{equation}
\begin{aligned}
x^i(z)_\pm &= \sum_{n\gtrless 0} x^i_n z^{-n}, \\
x^*_{i}(z)_\pm &= \sum_{n \gtrless 0} x^*_{i,n} z^{-n} + \frac{\log(z)}{2}  \varepsilon_{ijk} W^j x^k(z)_\pm, \\
\end{aligned}
\label{eq:2.30}
\end{equation}
Using \eqref{eq:2.20}--\eqref{eq:2.22}, the corresponding commutators are
\begin{equation}
\begin{aligned}
{[}x^*_{i}(z)_+, x^j(w)_-] &=  - \delta_i^j \sum_{n > 0} \frac{z^{-n}w^n}{n} = \delta_{i}^j i_{z,w} \log \left( 1 - \frac{w}{z} \right), \\
{[}x^i(z)_+, x^*_{j}(w)_-] &= \delta_i^j i_{z,w} \log\left(1 - \frac{w}{z} \right) ,\\
[x^*_i(z)_+, x^*_j(w)_-] &=  - \sum_{m,n >0}  \frac{m -n}{ 2 mn} \varepsilon_{ijk} x^k_{m-n} w^n z^{-m} + i_{z,w} \varepsilon_{ijk} W^k L \left( \frac{w}{z} \right).
\end{aligned}
\label{eq:2.31}
\end{equation}
The remaining commutators can be calculated using
\begin{equation}
\begin{aligned}
x^i(z) &= x^i(z)_+ + x^i(z)_- + x^i_0 + \log(z) W^i, \\
x^*_i(z) &= x^*_i(z)_+ + x^*_{i}(z)_- + x^*_{i,0} + \log(z) Q_i, \\
	Q_i &:= P_i - \frac{1}{2} \varepsilon_{ijk} x^j_0 W^k. 
\end{aligned}
\label{eq:2.31b}
\end{equation}
It follows from \eqref{eq:2.27} that $Q_i$ acts on functions as $\partial_{x^i}$ i.e.\
\begin{equation}
\begin{aligned}
{[}Q_i, x^*_j(w)_\pm] &= - \frac{1}{2} \varepsilon_{ijk} x^k(w)_\pm, \\
{[}x^*_i(z)_\pm, x^*_{j,0}] &= \frac{1}{2} \varepsilon_{ijk}\left[ \sum_{n \gtrless 0} x^k_n \frac{z^{-n}}{n} + \log(z) x^k(z)_\pm \right].
\label{eq:2.31c}
\end{aligned}
\end{equation}
A straightforward computation shows 
\begin{multline}
[x^*_i(z), x^*_j(w)] = \varepsilon_{ijk} W^k \left[ \sum_{m\neq 0} \frac{w^m z^{-m}}{m^2} + \frac{1}{2} \log\left(	\frac{z}{w} \right) \sum_{m \neq 0} \frac{z^{-m}w^m}{m} \right] + \\ \frac{1}{2} \varepsilon_{ijk} \left( \log(z) - \log(w) \right) \left( \tilde{x}^k(z) - \tilde{x}^k(w) \right) + \frac{1}{2} \varepsilon_{ijk} \sum_{n \neq 0} x^k_n \frac{w^{-n}}{n} +  \\ \frac{1}{2} \varepsilon_{ijk} \sum_{n \neq 0} x^k_n \frac{z^{-n}}{n} +  \varepsilon_{ijk} \sum_{m,n \neq 0} \frac{m+n}{2mn} x^k_{m+n} z^{-m} w^{-n} 
\label{eq:2.32}
\end{multline}
where $\tilde{x}^k$ is defined as in \eqref{eq:2.12}. Tn the first line is nothing but the sum of expansion of $L(w/z)$ in the domain $i_{z,w}$ with  the expansion of $L(z/w)$ in the domain $i_{w,z}$. Since
\begin{multline}
- \sum_{n \neq 0} \frac{w^n z^{-n}}{n} \left( \sum_{m \in \mathbb{Z}} x^k_m z^{-m} - x^k_m w^{-m} \right) =  \sum_{\stackrel{n \neq 0}{m \in \mathbb{Z}}} x^k_m \frac{w^{n-m} z^{-n}}{n} +\\ \sum_{\stackrel{n \neq 0}{m \in \mathbb{Z}}} x^k_m \frac{w^{-n} z^{n-m}}{n} =  \sum_{\stackrel{n \neq 0}{m \in \mathbb{Z}}} x^k_{m+n} \frac{w^{-m}z^{-n}}{n} +  \sum_{\stackrel{n \neq 0}{m \in \mathbb{Z}}} x^k_{m+n} \frac{w^{-n}z^{-m}}{n} = \\ \sum_{m,n \neq 0} x^k_{m+n} w^{-m} z^{-n} \left( \frac{1}{n} + \frac{1}{m} \right) + \sum_{n \neq 0} x^k_n \frac{z^{-n}}{n} + \sum_{n \neq 0} x^k_n \frac{w^{-n}}{n}. 
\label{eq:2.33}
\end{multline}
Substitution in \eqref{eq:2.32} yields
\begin{multline}
[x^*_{i}(z), x^*_j(w)] = \varepsilon_{ijk} W^k \left( i_{z,w} L\left( \frac{w}{z}  \right) + i_{w,z} L\left( \frac{z}{w} \right) \right) + \\ \frac{1}{2} \varepsilon_{ijk} \left( i_{z,w} \log(z-w) - i_{w,z} \log(w-z) \right) \left( \tilde{x}^k(z) - \tilde{x}^k(w) \right),
\label{eq:2.34}
\end{multline}
which coincides with \eqref{eq:2.13}.

\section{$4$-point functions and the reflection identity} \label{sec:4point}
Armed with \eqref{eq:2.13}  (or equivalently with \eqref{eq:2.34}, \eqref{eq:2.13c} and \eqref{eq:2.31}) we can calculate $4$-point HDFT correlators of observables corresponding to vacuum states. Ideally, one would would like to compute arbitrary expressions of the form 
\begin{equation}
\langle k| :f(z): :g(w): h \rangle, 
\label{eq:3.1}
\end{equation}
for $f,g,h$, and $k$ in $L^2(Y)$ expressed in terms of a suitable basis of $L^2(Y)$. 

Since there are operators in our theory acting on functions by differentiation it would be natural to choose a basis of smooth functions. While such a basis of $L^2(Y)$ can be written using Jacobi theta functions, we postpone the calculation of the corresponding HDFT correlators to future work. The reader is referred to appendix \ref{appendix} for an elementary discussion of the relevant harmonic analysis. For simplicity, in the present paper we focus on the natural exponential Fourier basis of $L^2(Y)$.

For each tuple $\alpha=(w^i, p_i) \in \tfrac{1}{2 \pi \sqrt{-1}} \cdot \Gamma$, we may consider the function 
\begin{equation}
e^\alpha = \exp( w^i x^*_i + p_i x^i ) \in L^2(Y),
\label{eq:3.2}
\end{equation}
where $x^i$ and $x^*_i$ are defined on a fundamental domain for $\Gamma \backslash G$ so that this function is in $L^2(Y)$ but is not continuous. Nevertheless, we can formally construct the corresponding  vertex operator. Defining
\begin{equation}
\alpha(z)_\pm = w^i x^*_i(z)_\pm + p_i x^i(z)_\pm, \quad \alpha_0 = w^i x^*_{i,0} + p_i x^i_0, \quad \alpha_l = w^i Q_i + p_i W^i,
\label{eq:3.2b}
\end{equation}
we set
\begin{equation}
e^\alpha (z) = :\exp(\alpha(z)):   \: := e^{\alpha_0} z^{\alpha_l} \exp (\alpha(z)_-) \exp (\alpha(z)_+).
\label{eq:3.3}
\end{equation}
where normal ordering is dictated by \eqref{eq:2.30}. 

Similar 4-point correlators of exponentials arise in the study of lattice vertex algebra where they can be evalue as $(z-w)^{(\alpha,\beta)}$ where $(,)$ is given by the natural pairing of the lattice $\Gamma$. To see what happens in presence of $H$-flux, let $\alpha = (w^i_\alpha, p_{i,\alpha})$ be a tuple in $\tfrac{1}{2 \pi \sqrt{-1}} \Gamma$ and similarly for $\beta$ and $\gamma$. If $e^\delta$ denotes the product of the functions $e^\alpha, e^\beta$ and $e^\gamma$, then 
\begin{multline}
\langle e^\delta| e^\alpha(z) e^\beta(w) e^\gamma \rangle = \\ \langle e^\delta| e^{\alpha_0} z^{\alpha_l} \exp(\alpha(z)_-) \exp(\alpha(z)_+) e^{\beta_0} w^{\beta_l} \exp(\beta(w)_-) \exp(\beta(w)_+) e^\gamma \rangle,
\label{3.4}
\end{multline}
 Since,
 \begin{equation}
\begin{aligned}
\langle e^{\delta} | e^{\alpha_0} = \langle e^{- \alpha} \cdot e^\delta |, && \langle e^\delta | z^{\alpha_l} = z^{(\alpha,\delta)} \langle e^\delta |, \\    \langle e^\delta| \exp(\alpha(z)_-) = \langle e^\delta |,\qquad &&   \exp(\alpha(z)_+) | e^\gamma \rangle = |e^\gamma \rangle
\end{aligned}
\label{eq:3.6}
\end{equation}
we only need to commute the annihilation part of $\alpha(z)$ past the creation part of $\beta(w)$ i.e.\
\begin{equation}
z^{(\alpha|\delta)} w^{(\beta,\delta- \alpha)} \langle e^{-\beta} \cdot e^{-\alpha} \cdot e^\delta| [\exp(\alpha(z)_+), e^{\beta_0} w^{\beta_l} \exp(\beta(w)_-)] | e^\gamma \rangle
\label{eq:3.8}
\end{equation}
We need to compute
\begin{equation}
\begin{aligned}
{[\alpha}(z)_+, \beta_0] &= \frac{1}{2} \varepsilon_{ijk} w^i_\alpha w^j_\beta \left[ \sum_{n > 0} x^k_n \frac{z^{-n}}{n} + \log(z) x^k(z)_+ \right],  \\
[\alpha(z)_+, \beta_l] &= - \frac{1}{2} \varepsilon_{ijk} w^i_\alpha w^j_\beta x^k(z)_+ \\
{[}\alpha(z)_+, \beta(w)_-] &= i_{z,w} \left[ (\alpha,\beta) \log\left( 1-\frac{w}{z} \right)  + \varepsilon_{ijk} w^i_\alpha w^j_\beta W^k L\left( \frac{w}{z} \right) \right] - \\ & \quad \varepsilon_{ijk} w^i_\alpha w^j_\beta \sum_{m,n >0} \frac{m-n}{2 mn} x^k_{m-n} w^{n} z^{-m}.
\end{aligned}
\label{eq:3.8b}
\end{equation}
For later convenience, we separate annihilation and creation contributions in the last commutator of \eqref{eq:3.8b}
\begin{multline}
[\alpha(z)_+, \beta(w)_- ] = i_{z,w} \left( (\alpha,\beta) \log\left( 1-\frac{w}{z} \right)  + \varepsilon_{ijk} w^i_\alpha w^j_\beta W^k L\left( \frac{w}{z} \right) \right) + \\ \varepsilon_{ijk} w^i_\alpha w^j_\beta   \sum_{n > 0} \sum_{m = 1}^{n-1} \left(\frac{m}{2n(n-m)} x^k_{-m} w^n z^{m-n} - \frac{m}{2(n-m) n} x^k_m w^{n-m} z^{-n}\right).
\label{eq:3.8c}
\end{multline}
Since in the commutator all annihilators  and creators will kill either $|e^\gamma\rangle$ or $\langle e^\delta|$ only their zero modes are relevant for this computation. This allows us to conclude that
\begin{multline}
\langle e^\delta | e^\alpha(z) e^\beta(w) | e^\gamma \rangle = z^{(\alpha,\delta)} w^{(\beta,\delta- \alpha)} \times \\ i_{z,w} \exp\left( (\alpha,\beta) \log\left( 1-\frac{w}{z} \right) + \varepsilon(\alpha,\beta,\gamma) L\left( \frac{w}{z} \right) \right) \langle e^{-\beta} e^{-\alpha} e^{\delta}| e^\gamma \rangle = \\
i_{z,w} z^{(\alpha,\alpha) + (\alpha,\gamma)} w^{(\beta,\beta)+(\beta,\gamma)} (z-w)^{(\alpha,\beta)} \exp\left( \varepsilon(\alpha,\beta,\gamma) L\left( \frac{w}{z} \right) \right)
\label{eq:3.10}
\end{multline}
where $\varepsilon(\alpha,\beta,\gamma) = \varepsilon_{ijk} w^i_\alpha w^j_\beta w^k_\gamma$. 
We recognize in (3.10) rational factors which are reminiscent of the analogue calculation for lattice vertex algebras. Commutation of $\alpha$ and $\beta$ affects the calculation by a sign $(-1)^{(\alpha,\beta)}$. This can be accounted for in analogy with the lattice vertex algebra i.e.\ by either introducing fermionic fields or by adding a $2$-cocycle in the definition of the fields. However, the recurring analogy with lattice vertex algebra cannot be pushed any further because\eqref{eq:3.10} also factors involving dilogarithms, which can be thought of the effect of having introduced a $3$-cocycle $\varepsilon$. Since dilogarithmic factors appear even when \emph{momenta} $p_{\alpha,i} = 0$, let us assume $\alpha = (w^i_\alpha, 0) \in \tfrac{1}{2\pi \sqrt{-1}} \Gamma$ (and similarly for $\beta$ and $\gamma$). We see that the $4$-point function $\langle e^\delta| e^\alpha(z) e^\beta(w)| e^\gamma \rangle$ converges in the domain $|w/z| < 1$ and it gives rise to a flat section of the bundle $\mathcal{L}_{(1-\frac{w}{z},\frac{w}{z})}$ on $\mathbb{CP}^1 \setminus \{0,1,\infty\}$ (with coordinate $\tfrac{w}{z}$). Commuting $\alpha$ with $\beta$ and noting that $\varepsilon$ is skew-symmetric, we obtain
\begin{equation}
\langle e^\delta| e^\beta(w) e^\alpha(z) |e^\gamma \rangle =i_{w,z} \exp\left( - \varepsilon(\alpha,\beta,\gamma) L\left( \frac{z}{w} \right) \right)
\label{eq:3.11}
\end{equation}
which converges to the expansion of a flat section of the bundle $\mathcal{L}^{-1}_{(1-\frac{z}{w},\frac{w}{z})}$ in the domain $|w/z| > 1$. Moreover, these sections are identified under the natural isomorphism \eqref{eq:1.30c} given by the bimultiplicativity property of Deligne's construction. 

We conclude this section by offering a geometrical interpretation of the HDFT correlators that we just computed.
\begin{prop}
Let $w_i \in \mathbb{Z}^3$, $i=1,\dots 3$ and denote by $e^{w_i} \in L^2(Y)$ be the corresponding vacuum states\footnote{Here we abuse notation, we should write $\exp(\sum_j w_i^j x^*_j)$ instead.}. Consider an Euclidean tetrahedron $\Delta_E$ in $\mathbb{R}^3$ with vertices given by $0$ and $w_i$.  Consider a ideal tetrahedron $\Delta_H$ with vertices at $\infty,z,w,0\in \mathbb{CP}^1\cong \partial\mathbb H_3$. Then \footnote{Here we abuse notation since the volume of $\Delta_H$ is only the imaginary part of the dilogarithm, therefore the phase of the correlator is given by the RHS of \eqref{eq:3.12}.} for any $\psi\in L^2(Y)$
\begin{equation}
\langle \psi| e^{w_1}(z) e^{w_2}(w) | e^{w_3} \rangle = i_{z,w} \exp\bigl( \vol (\Delta_E) \vol (\Delta_H) \bigr) \langle \psi | e^{w_1} e^{w_2} e^{w_3} \rangle
\label{eq:3.12}
\end{equation}
\label{prop:1}
\end{prop}
A consequence of our calculations in section \ref{section5}, will be this statement can generalized to arbitrary ideal tetrahedra i.e.\ to the arbitrary insertion points in $\mathbb CP^1$. 

\section{$5$-point functions and the pentagonal identity}\label{section5} In spite of having deduced in two distinct ways that the appearanceof dilogarithmic singularities is unavoidable, one might still wonder to what extent dilogarithms are intrinsic feature of  HDFT or rather something that pertains the particular examples under consideration. We believe that the calculation of 5-point functions provides evidence for the former conclusion. More precisely, we show that the highly non-trivial pentagonal identity \eqref{eq:1.26} is intimately related to the factorization structure of HDFT correlators. 

As in section \ref{sec:4point}, we restrict to the calculation of correlators of $5$ vacuum states corresponding to exponential functions. Moreover, we focus on those exponentials having fiber coordinates $x^*_i$ for argument, i.e.\ (taking the point of view of $\tilde X$) to those states with non-trivial winding but vanishing momenta. Just as before, let $\alpha = (w^i_\alpha) \in \mathbb{Z}^3$ and similarly consider vectors $\beta, \gamma$ and $\delta$. Setting
\begin{equation}\nonumber
e^\psi:=e^\alpha e^\beta e^\gamma e^\delta
\end{equation}
we focus on the evaluation the correlator
\begin{equation}
\langle e^\psi| e^\alpha(z) e^\beta(w) e^\gamma(t) | e^\delta \rangle.
\label{eq:4.1}
\end{equation}
which unfolds along the following lines. First of all, let us point out that our task is somewhat simplified owing to the fact that ${\rm span}\{\alpha,\beta,\gamma,\delta\}$ is a subgroup of $\Gamma$ which is isotropic with respect to the tautological pairing. 
Let us sketch the computation. Se begin by applying the annihilation part of $\gamma$ to $|e^\delta\rangle$ and the creation part of $\alpha$ to $\langle e^\psi|$. This transforms $\langle e^\psi |$ into $\langle e^{-\alpha} e^\psi|$. Next we commute the annihilation part of $\beta$ past the creation part of $\gamma$. Commuting it past $\gamma_0$ produces a term (cf. \eqref{eq:3.8b}) 
\begin{equation}
\exp \left( \frac{1}{2}	\varepsilon_{ijk} w^i_\beta w^j_\gamma \left[ \sum_{n > 0} x^k_n  \frac{w^{-n}}{n} + \log(w) x^k(w)_+ \right] \right).
\label{eq:4.2}
\end{equation}
This is an annihilation term so it will produce nothing when applied to $|e^\delta\rangle$. Similarly, commuting past $\gamma_l$ yields a further annhilation term that can be directly applied to $|e^\delta\rangle$. The only creation part comes from \eqref{eq:3.8c} and the winding operator gives us a factor:
\begin{equation}
\exp \left(\varepsilon(\beta,\gamma,\delta) L\left( \frac{t}{w} \right) \right).
\label{eq:4.3}
\end{equation}
The creation part from \eqref{eq:3.8c} is given by
\begin{equation}
\varepsilon_{ijk} w^i_\beta w^j_\gamma \sum_{n > 0} \sum_{m=1}^{n-1} \frac{m}{2n(n-m)} x^k_{-m} t^n w^{m-n}
\label{eq:4.4}
\end{equation}
which needs to be commuted past the annihilation part of $\alpha(z)$. This in turn produces a factor
\begin{equation}
\exp\left(- \frac{1}{2} \varepsilon(\alpha,\beta,\gamma) \sum_{n>0} \sum_{m=1}^{n-1} \frac{z^{-m} w^{m-n} t^n}{n(n-m)} \right)
\label{eq:4.5}
\end{equation}
At this point we have the factors \eqref{eq:4.3} and \eqref{eq:4.5} where $\beta$ and $\gamma$ are already in normal order. We reduced to computing
\begin{equation}
\langle e^{-\alpha} e^\psi| e^{\alpha(z)_+} e^{\beta_0} w^{\beta_l} e^{\beta(w)_-} e^{\gamma_0} t^{\gamma_l} e^{\gamma(t)_-} | e^{\delta} \rangle\, .
\label{eq:4.6}
\end{equation}
Commuting $\alpha(z)_+$ past $\beta_0$ produces an annihilation term  (as in the first equation of \eqref{eq:3.8b}) which needs to be commuted past the creation part of $\gamma$. A straightforward computation shows that this gives another factor
\begin{equation}
\exp \left[  \frac{1}{2} \varepsilon(\alpha,\beta,\gamma) \left(\log(z) \log\left( 1-\frac{t}{z} \right) - \mathrm{Li}_2 \left( \frac{t}{z} \right) \right) \right] 
\label{eq:4.7}
\end{equation}
Commuting $\alpha(z)_+$ past $\beta_l$ produces a further annihilation term (as in the second equation of \eqref{eq:3.8b}) which, when commuted past $\gamma$ yields a factor
\begin{equation}
\exp\left[- \frac{1}{2} \varepsilon(\alpha,\beta,\gamma) \log(w) \log\left( 1 - \frac{t}{z} \right)  \right].
\label{eq:4.8}
\end{equation}
Finally the annihilation term produced by $[\alpha(z)_+, \beta(w)_-]$ needs to be commuted past $\gamma$ and this gives rise to:
\begin{equation}
\exp\left[ \frac{1}{2} \varepsilon(\alpha,\beta,\gamma) \sum_{n > 0} \sum_{m=1}^{n-1} \frac{z^{-n} w^{n-m} t^m}{(n-m)n}  \right].
\label{eq:4.9}
\end{equation}
The corresponding term with a winding operator from the commutator $[\alpha(z)_+, \beta(w)_-]$ can be either applied to $\langle e^{-\beta} e^{-\alpha} e^\psi|$ or can be commuted past $\gamma_0$ and applied directly to $|e^\delta\rangle$ to obtain
\begin{equation}
\exp\left( \varepsilon(\alpha,\beta,\gamma) L\left( \frac{w}{z}  \right) + \varepsilon(\alpha,\beta,\delta) L\left( \frac{w}{z} \right) \right).
\label{eq:4.10}
\end{equation}
We are left to compute
\begin{equation}
\langle e^{-\beta} e^{-\alpha} e^\psi| e^{\alpha(z)_+} e^{\gamma_0} t^{\gamma_l} e^{\gamma(t)_-} | e^\delta \rangle.
\label{eq:4.11}
\end{equation}
As before, commuting $\alpha(z)_+$ past $\gamma_0$ and $\gamma_l$ produces annihilation terms that fixes $|e^\delta\rangle$. The only relevant term is now the winding operator from \eqref{eq:3.8c} which is responsible for
\begin{equation}
\exp\left( \varepsilon(\alpha,\gamma,\delta) L\left( \frac{t}{z} \right) \right).
\label{eq:4.12}
\end{equation}
Collecting together the above calculations, conclude that $5$-point function is given as the expansion in the domain $|z| > |w| > |t|$ of the exponential of 
\begin{multline}
\varepsilon(\beta,\gamma,\delta) L\left( \frac{t}{w} \right) + \varepsilon(\alpha,\beta,\gamma) L\left( \frac{w}{z} \right) + \varepsilon(\alpha,\beta,\delta) L\left( \frac{w}{z} \right) + \varepsilon(\alpha,\gamma,\delta) L\left( \frac{t}{z} \right) - \\ \frac{1}{2} \varepsilon(\alpha,\beta,\gamma) \left( \mathrm{Li}_2 \left( \frac{t}{z} \right) + \log\left( \frac{w}{z} \right) \log\left( 1-\frac{t}{z} \right) \right) + \\ \frac{1}{2} \varepsilon(\alpha,\beta,\gamma) \sum_{n > 0} \sum_{m=1}^{n-1} \frac{z^{-n} w^{n-m} t^m}{(n-m)n} - \frac{1}{2} \varepsilon(\alpha,\beta,\gamma) \sum_{n>0} \sum_{m=1}^{n-1} \frac{z^{-m} w^{m-n} t^n}{n(n-m)} 
\label{eq:4.13}
\end{multline}
In particular, if $\delta=0$ and $t=0$ we recover \eqref{eq:3.10}.

The role of the pentagonal identity emerges from a careful analysis of the last two terms in \eqref{eq:4.13}. To this end, we need to collect some information regarding the expansion of   dilogarithms on various domains. We denote by $i_{z,w,t}$ the expansion of a function of $z,w$ and $t$ in the domain $|z| \gg |w| \gg |t|$. In particular 
\begin{multline}
i_{z,w,t} \left[ \mathrm{Li}_2\left( \frac{t-w}{t-z} \right) - \mathrm{Li}_2 \left( \frac{t}{t-z} \right) \right] = \sum_{n > 0} \sum_{j = 1}^n \frac{(-1)^{n+j}}{n^2} \binom{n}{j} \frac{t^{n-j} z^{-n} w^j}{\left( 1-\frac{t}{z} \right)^n} \\ = \sum_{n > 0} \sum_{j=1}^n \sum_{m \geq 0} \frac{(-1)^{n+j}}{n^2} \binom{n}{j}\binom{n+m-1}{n-1} t^{m+n-j} z^{-m-n} w^j \\ = \sum_{k > 0} \sum_{j = 1}^{k} t^{k-j} z^{-k} w^j \sum_{n=j}^k \frac{(-1)^{n+j}}{n^2} \binom{n}{j} \binom{k-1}{n-1} \\ = \sum_{k > 0} \sum_{j = 1}^{k} (-1)^j \frac{(k-1)!}{j! (k-j)!} t^{k-j} z^{-k} w^j \sum_{n=j}^k \frac{(-1)^n}{n} \frac{(k-j)!}{(n-j)! (k-n)!} \\ = \sum_{k>0} \sum_{j=1}^{k} \frac{(-1)^j}{k} \binom{k}{j} t^{k-j} z^{-k} w^j \sum_{n=j}^k \frac{(-1)^n}{n} \binom{k-j}{n-j}.
\label{eq:binom3}
\end{multline}
The inner sum can be further evaluated as follows. For $(j>0)$
\begin{equation}
\int_0^1 x^{j-1}(1-x)^{k} dx = \int_0^1 \sum_{n=0}^k (-1)^n \binom{k}{n} x^{n+j-1} dx  =  \sum_{n=0}^k \frac{(-1)^n}{n+j} \binom{k}{n}
\label{eq:binom1}
\end{equation}
On the other hand 
\begin{equation}
\int_0^1 x^{j-1} (1-x)^k dx = \beta(j,k+1) := \frac{\Gamma(j) \Gamma(k+1)}{\Gamma(j+k+1)} = \frac{(j-1)! k!}{(j+k)!} = \frac{1}{j \binom{j+k}{j}}.
\label{eq:binom2}
\end{equation}
so that, combining \eqref{eq:binom1}, \eqref{eq:binom2} and  \eqref{eq:binom3} we obtain:
\begin{equation}
i_{z,w,t} \left[ \mathrm{Li}_2\left( \frac{t-w}{t-z} \right) - \mathrm{Li}_2 \left( \frac{t}{t-z} \right) \right] = \sum_{k>0} \sum_{j=1}^k \frac{1}{k j} t^{k-j}z^{-k} w^{j},
\end{equation}
or equivalently
\begin{multline}
i_{z,w,t} \left[ \mathrm{Li}_2\left( \frac{t-w}{t-z} \right) - \mathrm{Li}_2 \left( \frac{t}{t-z} \right) - \mathrm{Li}_2 \left( \frac{w}{z} \right) \right] = \\ \sum_{n>0} \sum_{m=1}^{n-1} \frac{1}{n m} t^{n-m}z^{-n} w^{m}.
\label{eq:lamiaexpansion}
\end{multline}
We recognize the RHS of \eqref{eq:lamiaexpansion} as the first of the sums in \eqref{eq:4.13} while the last sum in \eqref{eq:4.13} arises by sending $z \mapsto t^{-1}, w \mapsto w^{-1}$ and $t \mapsto z^{-1}$ we recognize the last sum in \eqref{eq:4.13} (note also that the domain of expansion is invariant as $i_{z,w,t} = i_{t^{-1},w^{-1},z^{-1}}$).

Putting together all of the above, we obtain that our $5$-point HDFT correlator is the expansion in the domain $i_{z,w,t}$ of the exponential of
\begin{multline}
\varepsilon(\beta,\gamma,\delta) L\left( \frac{t}{w} \right) + \varepsilon(\alpha,\beta,\gamma) L\left( \frac{w}{z} \right) + \varepsilon(\alpha,\beta,\delta) L\left( \frac{w}{z} \right) + \varepsilon(\alpha,\gamma,\delta) L\left( \frac{t}{z} \right) - \\ \frac{1}{2} \varepsilon(\alpha,\beta,\gamma) \left( \mathrm{Li}_2 \left( \frac{t}{z} \right) + \log\left( \frac{w}{z} \right) \log\left( 1-\frac{t}{z} \right) \right) + \\ \frac{1}{2} \varepsilon(\alpha,\beta,\gamma)  
 \left[ \mathrm{Li}_2\left( \frac{t-w}{t-z} \right) - \mathrm{Li}_2 \left( \frac{t}{t-z} \right) - \mathrm{Li}_2 \left( \frac{w}{z} \right) \right] - \\
	\frac{1}{2} \varepsilon(\alpha,\beta,\gamma) \left[ 
\mathrm{Li}_2\left( \frac{z^{-1}-w^{-1}}{z^{-1}-t^{-1}} \right) - \mathrm{Li}_2 \left( \frac{t}{t-z} \right) - \mathrm{Li}_2 \left( \frac{t}{w} \right) \right],
\label{eq:4.18}
\end{multline}
which further simplifies to 
\begin{multline}
\varepsilon(\beta,\gamma,\delta) L\left( \frac{t}{w} \right) + \varepsilon(\alpha,\beta,\delta) L\left( \frac{w}{z} \right) + \varepsilon(\alpha,\gamma,\delta) L\left( \frac{t}{z} \right) + \\ \frac{1}{2} \varepsilon(\alpha,\beta,\gamma) \left( L\left( \frac{w}{z} \right) - L\left( \frac{t}{z} \right) + L \left( \frac{w-t}{z-t} \right) - \right. \\ \left.L \left( \frac{w^{-1} - z^{-1}}{t^{-1} - z^{-1}} \right) +  L \left( \frac{t}{w} \right) \right).
	\label{eq:4.19}
\end{multline}
Using the pentagonal identity \eqref{eq:1.26}, we finally arrive to the following form of the $5$-point function
\begin{multline}
\langle e^\psi | e^\alpha(z) e^\beta(w) e^\gamma(t) | e^\delta \rangle = i_{z,w,t} \exp \left( \varepsilon(\beta, \gamma, \delta) L\left( \frac{t}{w}  \right)  + \right. \\ \left.   \varepsilon(\alpha,\beta,\delta) L\left( \frac{w}{z} \right) + \varepsilon(\alpha,\gamma,\delta) L\left( \frac{t}{z} \right) + \varepsilon(\alpha,\beta,\gamma) L\left( \frac{w-t}{z-t} \right)  \right).
\label{eq:4.22}
\end{multline}
\begin{rem}
Note that letting $\delta=0$ in \eqref{eq:4.22} yields 
\begin{equation}
\langle e^\psi | e^\alpha(z) e^\beta(w) e^\gamma(t) | 0 \rangle = i_{z,w,t} \exp\left( \varepsilon(\alpha,\beta,\gamma) L\left( \frac{w-t}{z-t} \right) \right).
\label{eq:4.24}
\end{equation}
This should be thought of as generalization of proposition \ref{prop:1} with insertion points at $\infty$, $z$, $w$ and $t$ respectively. Notice that the section in the RHS of \eqref{eq:4.24} corresponds to 
\[ \langle e^\psi | e^\alpha(z-t) e^\beta(w-t) | e^\gamma \rangle. \]
In a vertex algebra, the equality of this expression with the the LHS of \eqref{eq:4.24} would be a consequence of associativity which in turn is related to the factorization property of the correlation functions. 

Similarly,  if $\langle \psi | = \langle 0 |$ the corresponding function is not zero only when $e^\delta = e^{-\gamma} e^{-\beta} e^{-\alpha}$. Using the pentagonal identity, we obtain 
\begin{equation}
\langle 0| e^\alpha(z) e^\beta(w) e^\gamma(t) |e^\delta \rangle = i_{z,w,t} \exp\left( \varepsilon(\alpha,\beta,\gamma) L\left( \frac{w^{-1} - z^{-1}}{t^{-1}-z^{-1}} \right) \right).
\label{eq:4.22b}
\end{equation}
\label{rem:2}
which may be viewed as a consequence of Proposition \ref{prop:1} where the four points are placed at $z$, $w$, $t$, and $0$ respectively. 
\end{rem}
Equation \eqref{eq:4.22} therefore implies the following factorization property for this type of $5$-point functions  
\begin{multline}
\langle e^\psi | e^\alpha(z) e^{\beta}(w) e^\gamma(t) | e^\delta \rangle = \langle e^{-\alpha} e^\psi | e^{\beta}(w) e^\gamma(t) | e^\delta \rangle  \times \\ \langle e^{-\gamma} e^\psi | e^{\alpha}(z) e^\beta(w) | e^\delta \rangle \times \langle e^{-\beta} e^\psi | e^{\alpha}(z) e^{\gamma}(t) | e^\delta \rangle \times \\ \langle e^{-\delta} e^\psi | e^\alpha(z-t) e^\beta(w-t) |e^\gamma \rangle.
\label{eq:4.23}
\end{multline}
In terms of volumes of ideal tetrahedra, this can be restated as follows:
\begin{prop}
Let $w^i \in \mathbb{Z}^3$, $i=1,\dots,4$, let $e^{w_i} \in L^2(Y) \subset \mathcal{H}$ the corresponding vacuum states and let $\Delta_{E,i}$ be the Euclidean tetrahedron with vertices at $0$ and $w_1, \dots, \widehat{w_i}, \dots, w_4$. Moreover, let $z_4=0$ and choose in addition $z_1,\dots, z_3\in \mathbb{CP}^1$. Finally for $i=1,\ldots,4$ let $\Delta_{H,i}$ be the ideal tetrahedron in $\mathbb H_3$ with vertices at $\infty,\ldots,\hat z_i,\ldots,z_4\in \mathbb{CP}^1$. Then for any $\psi \in L^2 (Y)$
\begin{multline}
\langle \psi (\infty) | e^{w_1}(z_1) e^{w_2}(z_2) e^{w_3}(z_3) |e^{w_4}(0) \rangle = \\ i_{z,w,t} \prod_{i=1}^4 \langle \psi,  e^{w_1} \dots \widehat{e^{w_i}} \dots e^{w_4} \rangle  \exp \left(  \vol(\Delta_{E,i}) \vol(\Delta_{H,i}) \right).
\label{eq:4.25}
\end{multline}
\label{prop:2}
\end{prop}
\section{Hamiltonians and the equations of motion} \label{sec:5} 
We described the algebraic structure that HDFT imposes on the space of states \eqref{eq:1.19} and argued that HDFT correlators should be interpreted as sections of line bundles on a punctured $\mathbb{CP}^1$ minus some points. In this section we continue along the path outlined in the introduction and investigate the dynamics of our HDFT model. As described in the introduction, this is achieved by choosing a suitable Hamiltonian object in \eqref{eq:1.19} and impose \eqref{eq:1.11} as the fundamental equation of motion of the theory. 

Naively, one might be tempted to proceed as follows. Since the $6$-dimensional Lie algebra $\fg$ has a non-degenerate invariant form $(\cdot,\cdot)$ we have the corresponding embedding of the Virasoro algebra with central charge $6$ in the basis of \eqref{eq:2.4} by
\begin{equation}
H'(z) =  :\alpha^i(z) \beta_i(z):.
\label{eq:5.1}
\end{equation}
A straightforward computation shows
\begin{equation}
H'(z) \cdot H'(w) \sim \frac{\partial_w H'(w)}{z-w} + \frac{2 H'(w)}{(z-w)^2} + \frac{6/12}{(z-w)^4}
\label{eq:5.2}
\end{equation}
so that $H$ is a Virasoro field. However, the zero mode $H'_0$ of \eqref{eq:5.1} is the infinitesimal generator $\partial_\sigma$ of translations along the string parameter, and not $\tau$ as we would like. 

The point is that to prescribe dynamics one needs to decorate the target manifold with extra geometrical data. Concretely, fix the target to be either the Heisenberg nilmanifold $X$ or the twisted torus $\tilde X$ and pick on it a Riemannian metric $g_{ij}$ with inverse $g^{ij}$. Since HDFT contains currents associated with differential forms and vector fields, it makes sense to the define the field 
\begin{equation}
H = \frac{1}{2} g_{ij} dx^i dx^j + \frac{1}{2} g^{ij} \partial_{x^i} \partial_{x^j}.
\label{eq:5.a}
\end{equation}
and to take its zero-mode as the Hamiltonian of the theory.
\begin{rem}
While HDFT contains all fields associated with sections of the standard Courant algebroid $T \oplus T^*$ on the target manifold (possibly twisted by a gerbe), it also contains more general multi-valued fields due to the presence of logarithms and ``winding'' operators in the loop expansions. However, restriction to contractible loops makes winding operators irrelevant and one recovers the bosonic part of the \emph{chiral de Rham complex} (CDR) of \cite{malikov} (cf. \cite{bressler2} and \cite{alekseev} for the twisted case). Since the fields associated with sections of $T$ and $T^*$ do not include the conjugate variables to the winding operators, the computations in this section remain valid in the context of vertex algebras as computations in CDR. This can be interpreted as a consequence of the fact that the equations of motion satisfied by the scalar fields are independent of the winding. 
\label{rem:5}
\end{rem}
\subsection{The twisted torus}
For the twisted torus $\tilde{X} = T^3$ with coordinates $x^i$ and H-flux $\varepsilon = d\mathrm{vol}$, we consider the flat metric $g_{ij} = \delta_{ij}$. In terms of the currents $\beta_i$, $\alpha_i$ associated with $\partial_{x_i}$ $dx^i$, respectively, \eqref{eq:5.a} reads
\begin{equation}
H(z) = \frac{1}{2} \left( \sum_i :\alpha^i(z) \alpha^i(z): + :\beta_i(z) \beta_i(z):\right).
\label{eq:5.4}
\end{equation}
 In order to compute the equations of motion \eqref{eq:1.11} we need the OPE
\begin{equation}
\begin{aligned}
H(z) \cdot x^i(w) &\sim \frac{\beta_i(w)}{(z-w)} \\ H(z) \cdot \beta_i(w) &\sim \frac{\partial_w \alpha^i(w) - \varepsilon_{ijk} :\beta_j(w) \alpha^k(w):}{z-w} + \frac{\alpha^i(w)}{(z-w)^2}.
\label{eq:5.5}
\end{aligned}
\end{equation}
Let us pass to the coordinate $z=e^{2 \pi i \sigma}$. Since $\alpha^i$ is associated with $dx^i$, it follows that $\partial_\sigma x^i = \alpha^i$. We deduce the equations of motion
\begin{equation}
\partial_\tau x^i = \beta_i \qquad \partial_\tau \beta_i = \partial_\sigma \alpha^i - \varepsilon_{ijk} :\beta_j \partial_\sigma x^k:\,.
\label{eq:5.6}
\end{equation}
This is equivalent to the single second-order equation
\begin{equation}
\bigl( \partial_\tau^2 - \partial^2_\sigma \bigr) x^i  + \varepsilon_{ijk} \partial_\tau x^j \partial_\sigma x^k = 0.
\label{eq:5.7}
\end{equation}
which, for $\vec{x}\in\mathbb{R}^3$ can be rewritten as
\begin{equation}
\square^2 \vec{x} + \partial_\tau \vec{x} \times \partial_\sigma \vec{x} =0
\label{eq:5.8}
\end{equation}
where $\square^2 = (\partial_\tau^2 - \partial_\sigma^2)$ is the d'Alambertian. We impose  periodic boundary conditions
\begin{equation}
\vec{x}(\sigma + 1, \tau) = \vec{x}(\sigma, \tau) \mod \mathbb Z^3\, .
\label{eq:5.9}
\end{equation}
\subsection{The Heisenberg nilmanifold} We now turn to the $T$-dual model with target $X$, the Heisenberg nilmanifold defined in section \ref{sec:noncommut}. Given a Riemannian metric on $X$, the computation of the HDFT OPEs necessary to obtain the equations of motion is potentially non-trivial because derivatives of the metric (and thus the Christoffel symbols of the Levi-Civita connection) are involved. However, $X$ has a canonical metric $g_{ij}$ inherited from the left invariant measure on the Heisenberg group whose Christoffel symbols of the Levi-Civita connection are naturally identified with the structure constants of the Heisenberg Lie algebra. 

As explained in Section \ref{sec:noncommut} the current algebra \eqref{eq:1.18} associated with the standard Courant algebroid $T \oplus T^*$ on $X$ corresponds to the left invariant vector fields and differential forms \eqref{eq:1.14}. In terms of this basis and of coordinates $y^1, y^2,y^3$
\begin{equation}
ds^2 := g_{ij} dy^i dy^j = \sum_i Y^i Y^i,
\label{eq:5.10}
\end{equation}
whence
\begin{equation}
g_{ij} = \begin{pmatrix}
1 + \frac{1}{4} (y^3)^2 & \frac{1}{2} y^3 & - \frac{1}{4} y^1 y^3 \\ 
\frac{1}{2} y^3 & 1 & - \frac{1}{2} y^1 \\
- \frac{1}{4} y^1 y^3 & - \frac{1}{2} y^1 & 1 + \frac{1}{4} (y^1)^2
\end{pmatrix}.
\label{eq:5.11}
\end{equation}
with inverse tensor
\begin{equation}
g^{ij} = \begin{pmatrix}
1 & - \frac{y^3}{2} & 0 \\ 
- \frac{y^3}{2} & 1 + \frac{(y^1)^2 + (y^3)^2}{4} & \frac{y^1}{2} \\
		0& \frac{y^1}{2} & 1 \end{pmatrix}.
\label{eq:5.11b}
\end{equation}
The corresponding Hamiltonian can then be computed according to \eqref{eq:5.a}, with $y^i$ in place of $x^i$. In terms of \eqref{eq:1.14} this amounts to 
\begin{equation}
H = \frac{1}{2} \sum_i Y^i Y^i + \frac{1}{2} \sum_i X_i X_i
\label{eq:5.12}
\end{equation}
which which yields OPEs
\begin{equation}
\begin{aligned}
H(z) \cdot y^i(w) &\sim \frac{X^i(w)}{z-w} \\
H(z) \cdot X^1(w) &\sim \frac{\partial_w Y^1(w) + :Y^2(w) Y^3(w): - :X^3(w) X^2(w):}{z-w} + \frac{Y^2(w)}{(z-w)^2},\\
H(z) \cdot X^2(w) &\sim \frac{Y^2(w)}{z-w} + \frac{\partial_w Y^2(w)}{(z-w)^2} \\
H(z) \cdot X^3(w) &\sim \frac{\partial_w Y^3(w) + :X^1(w) X^2(w): - :Y^2(w)Y^3(w):}{z-w} + \frac{Y^3(w)}{(z-w)^2}
\end{aligned}
\label{eq:5.13}
\end{equation}
From the definition of the currents \eqref{eq:1.14} we have
\begin{equation}
\begin{gathered}
Y^1(z) = \partial_z y^1, \qquad Y^3(z) = \partial_z y^3, \\
		 Y^2(z) = \partial_z y^2 - \frac{1}{2} y^1 \partial_z y^3 + \frac{1}{2} y^3 \partial_z y^1\, . 
\label{eq:5.14}
\end{gathered}
\end{equation}
Replacing in the zero modes of \eqref{eq:5.13} and making the change of coordinates $z = e^{2\pi i \sigma}$ we arrive to the equations of motion for the scalar fields $y^i$ associated with our coordinates
\begin{equation}
\begin{aligned}
\square^2 y^1 &= \left( \partial_\sigma y^2 - \frac{1}{2} y^1 \partial_\sigma y^3 + \frac{1}{2} y^3 \partial_\sigma y^1 \right) \partial_\sigma y^3 - \partial_\tau y^3 \partial_\tau y^2 \\ 
\square^2 y^2 &= \frac{1}{2} y^3 \partial_\sigma y^2 - \frac{1}{2} y^1 \partial_\sigma y^3, \\
\square^2 y^3 &= \partial_\tau y^1 \partial_\tau y^2 - \left( \partial_\sigma y^2 - \frac{1}{2} y^1 \partial_\sigma y^3 + \frac{1}{2} y^3 \partial_\sigma y^1 \right) \partial_\sigma y^3
\end{aligned}
\label{eq:5.15}
\end{equation}
subject to the periodic boundary conditions:
\begin{equation}
\bigl( y^1(\sigma + 1), y^2(\sigma+1), y^3(\sigma + 1) \bigr) = \left( \gamma^1, \gamma^2, \gamma^3 \right) \cdot \bigl( y^1(\sigma), y^2(\sigma), y^3(\sigma) \bigr), 
\label{eq:5.16}
\end{equation}
for some $(\gamma^i)$ in the lattice $\Gamma \subset H(\mathbb{R})$. 

We are now in position to prove that T-duality is an isomorphism between the HDFT dynamical sigma models on $X$ and $\tilde X$.  According to DFT, we identify the coordinates $x^i$ of the twisted torus $\tilde{X}$ with the coordinates on the base of standard torus fibration of $Y$. Similarly, we identify the coordinates $y^1$ and $y^3$ of the Heisenberg nilmanifold $X$ with the coordinates $x^1$ and $x^3$ of $Y$ while $y^2$ identified with $-x^*_2$ as a coordinate \footnote{The minus sign here is due to the chosen group structure for the Heisenberg group, if we change the signs in \eqref{eq:1.14b} we would identify $y^2$ with $x^*_2$.}  on $Y$. The currents $X^i, Y^i$ of \eqref{eq:1.14} are identified with the currents $\beta_i, \alpha^i$ of section \ref{sec:noncommut}. Under these identifications, the Hamiltonian \eqref{eq:5.12} matches that of \eqref{eq:5.4}. This implies the {\it a priori} non-trivial equivalence of  the systems of equations \eqref{eq:5.8} and \eqref{eq:5.15}. 

\begin{rem}We remark here that the equivalence just described is a consequence of the defining equations \eqref{eq:2.7} (or \eqref{eq:5.14}). Consider for example the first equation of \eqref{eq:5.15} and identify $y^1 = x^1$, $y^3 = x^3$ and $y^2 = - x^*_2$.  From \eqref{eq:2.7} and  \eqref{eq:5.6}
\begin{equation}
\partial_\sigma y^2 - \frac{1}{2} y^1 \partial_\sigma y^3 + \frac{1}{2} y^3 \partial_\sigma y^1 = - \beta_2=\partial_\tau x^2\, .
\label{eq:5.17}
\end{equation}
Similarly, from the first equation of \eqref{eq:5.13}
\begin{equation}\nonumber
\partial_\tau y^2 = X^2=-\alpha^2=- \partial_\sigma x^2. 
\end{equation}
As a result, we can rewrite the first equation of \eqref{eq:5.15} as
\begin{equation}
\square^2 x^1 = - \partial_\tau x^2 \partial_\sigma x^3 + \partial_\tau x^3 \partial_\sigma x^2,
\label{eq:5.18}
\end{equation}
which coincides with the $i=1$ case of \eqref{eq:5.7}.
\label{rem:3}
\end{rem}
\section{Summary and Discussion}
\def\cH{\mathcal{H}}
In this article we considered the sigma model with target the Heisenberg nilmanifold $X = \Gamma \backslash H(\mathbb{R})$ and the $T$-dual sigma-model with target the twisted $3$-torus $\tilde{X}$.  Using HDFT, we constructed showed that Hilbert spaces of both theories are naturally identified with \eqref{eq:1.19}, which is constructed from the geometry of the double twisted torus $Y$. HDFT endows $\mathcal{H}$ with with an algebraic structure reminiscent of that of a vertex algebra.  In particular, given four vectors $\alpha$,$\beta$,$\gamma, \delta\in \cH$ it makes sense to compute the four-point HDFT correlators 
\begin{equation}
\langle \alpha| \beta(z) \gamma(w) |\delta \rangle, \qquad \langle \alpha| \gamma(w) \beta(z) | \delta \rangle\, .
\label{eq:6.1}
\end{equation}
These correlators turn out to be expansions in their respective domains of flat sections of certain natural line bundles on $\mathbb{CP}^1 \setminus \{0,1,\infty\}$ with connections. In HDFT, the notion of locality of vertex algebras is generalized to the existence of explicit isomorphisms between the corresponding line bundles which identify these sections.

As in any sigma-model with target a manifold $X$, one has well defined currents associated with vector fields (sections of $TX$) and differential forms (sections of $T^*X$). These fields satisfy an OPE that reflects the Courant-Dorfmann bracket and the natural pairing on $TX \oplus T^*X$. In our examples, $TX\oplus T^*X$ can be trivialized by a global orthogonal frame. In DFT language, $TX\oplus T^*X$ is traded for $TY$, the Courant-Dorfmann bracket for the Lie bracket on $TY$  and the corresponding currents can be thought of as generators of the affine current algebra \eqref{eq:1.18}. The scalar fields associated with the coordinate functions on $Y$ satisfy a simple OPE obtained by integrating \eqref{eq:1.18} and are subject to the relations \eqref{eq:2.6} and \eqref{eq:2.7}. Computing the singular part of this OPE \eqref{eq:2.13c} we stubble upon a novel phenomenon: the OPE contains dilogarithmic terms. 

One of the main goals of this paper is to point out how dilogarithmic singularities in our OPEs are an inevitable consequence of the simple integration process of \eqref{eq:1.18}. To give this observation the importance we believe it deserves, we limited our analysis to the calculation of $4$ and $5$ point functions associated with HDFT vacuum states. Using an idea of Deligne \cite{deligne}, we find that HDFT correlators are naturally actually sections of line bundles with connections on the worldsheet. Using the \emph{bimultiplicative} nature of Deligne's construction, HDFT correlators naturally reflect non-trivial functional equations satisfied by the dilogarithm. In the case of $4$ point functions, the reflection identity allows us to match the two sections in \eqref{eq:6.1}. 

Using the fact that the dilogarithm function measures the volume of ideal (hyperbolic) tetrahedra we see that the correlator of a collection of fields can be written interms of products of the volume of the Euclidean tetrahedra spanned by the winding vectors associated with those fields and the volume of the ideal tetrahedra defined by their insertion points on the $\mathbb{CP}^1$ worldsheet. For $5$-point functions this identification requires the pentagonal identity for the dilogarithm function. The group $\mathrm{SL}_2(\mathbb{Z})$ acts on the boundary $\mathbb{CP}^1$ of $\mathbb{H}_3$ preserving the volumes of these tetrahedra. The pentagonal identity ensures that our correlators are compatible with this action.

The construction of HDFT correlators is purely kinematic i.e.\ it doesn't require solving the equations of motion and the worldsheet coordinate $z$ is a formal parameter. Dynamics can be added to our HDFT models by prescribing Hamiltonian operators. We described the explicit $T$-duality isomorphism relating the HDFT algebra attached to $X$ to the HDFT algebra of $\tilde X$. Moreover, we proved that such an isomorphism intertwines the Hamiltonian associated with the flat metric on $\tilde X$ and the Hamiltonian associated with the left-invariant metric on $X$. This rigorously establishes $T$-duality between these two theories as full (i.e.\ containing all winding sectors and all oscillators) dynamical sigma models. This is generalizes an analogous result for tori \cite{kapustinorlov}.

Some open problems that will be addressed in future publications are 
\begin{enumerate}
\item While for simplicity we did not include fermions in our model, such an extension is not difficult and in fact rather natural. First of all, even dimensionality of the target can be achieved by adding add a $S^1$ factor to $X$ and the dual $(S^1)^*$ to $\tilde X$. According to a well known calculation, this establishes $X':=X\times S^1$  and $\tilde X':=\tilde X\times (S^1)^*$ as a mirror pair. For suitable choices of a symplectic structure of a symplectic structure on $X'$ and of a complex structure on $\tilde X'$. While topological reasons prevent both $X'$ and $\tilde X'$ (with H-flux) from being K\"ahler, these geometries are in fact generalized Calabi-Yau manifolds (but neither Calabi-Yau nor generalized Calabi-Yau metric manifolds). The formalism of \cite{heluanizabzine} applies in this situation producing an embedding of the $N=2$ superconformal vertex algebra in the corresponding Hilbert spaces. These two structures are intertwined by the natural isomorphism given by $T$-duality. This yields a statement of mirror-symmetry generalizing that of \cite{kapustinorlov} to our setup.

\item In the present paper we focused on correlators of exponential functions. Ultimately, one would like to understand correlators of actual smooth functions on the double twisted torus $Y$. 
Even though the expressions for these correlators turn out to be rather involved, we expect the structure to remain the same: the collection of $n$-point functions is given as flat sections of suitable line bundles with flat connections on the world-sheet, satisfying certain factorization relations. 

\item In a way, the HDFT algebra attached to $Y$ is very much reminiscent of that of the lattice vertex algebra. To illustrate this point, we observe how the construction of Deligne that we use \cite{deligne} is a global version of  the Contou-Carr\`ere symbol which is a key ingredient in the construction of the lattice chiral algebra of Belinson and Drinfeld \cite{beilinsondrinfeld}. Identifying the lattice $\Gamma$ in the definition of $Y$ with $\mathbb{Z}^6$ we constructed $\cH$ out of  the following data: $\Gamma$, a pairing $\Gamma \otimes \Gamma \rightarrow \mathbb{C}$, a $2$-cocycle on $\Gamma$ (see \eqref{eq:3.11} and the discussion following it) and a $3$-cocycle $\varepsilon$. It would be interesting to make precise the statement that the HDFT algebra of $Y$ is an analytic version of the lattice vertex algebra associated with a lattice twisted by a $3$-cocycle.  
 
\item The dilogarithm function has a natural quantization \cite{fadeev}. It is natural to ask if quantum dilogarithms also play a role in our construction. We conjecture this to be the case i.e.\ that the $4$-point functions of section \ref{sec:4point} evaluated on a ``non-commutative worldsheet'' with coordinates $z$ and $w$ satisfying $zw = q wz$ is an expression for the quantum dilogarithm. 
\end{enumerate}

\appendix

\section{The space of ground states}\label{appendix}
The goal of this appendix is to complement the kinematic description of our model with a geometric interpretation of the space $\mathcal H^0$ of ground states. In particular, we show that both the ground states of the twisted torus $\tilde X$ and those of the Heisenberg nilmanifold $X$ can be identified with functions on the double twisted torus Y. The algebraic structure behind this quantization map is rather subtle and we hope to return to this subject for a more comprehensive analysis in the near future. Here we restrict to a simple-minded but rather explicit treatment, in the spirit of this paper.

\subsection{Magnetic tori}\label{magnetic}
We begin with a quantum mechanical analogue: a particle moving on a magnetic  torus. Let  $T=\mathbb Z^2\backslash \mathbb R^2$ and let $x^1,x^2$ be coordinates on $\mathbb R^2$. For each $n\in \mathbb Z$, the form 
\begin{equation}\nonumber
A:=n x^1 dx^2\in \Omega^1(\mathbb R^2)
\end{equation}
descends to a $U(1)$-connection for a line bundle $\mathcal L$ on $T$. The space of states $\mathcal H$ of this system is a completion\footnote{Following \cite{kapustinorlov} we ignore completions and work with smooth functions (or their distributional limits) instead} of the space of smooth sections of $\mathcal L$
\begin{equation}\nonumber
C^\infty(T,A):=\{f\in C^\infty (\mathbb R^2) \,|\, f(x+\gamma)=f(x) {\rm Hol} (\gamma,x)\,,\, \forall \gamma\in \mathbb Z^2 \}\,, 
\end{equation}
where 
\begin{equation}\nonumber
{\rm Hol}(\gamma,x):=q^{\iota_\gamma A}\,.
\end{equation}
and
\begin{equation}\nonumber
q:=\exp(2\pi \sqrt{-1})\,.
\end{equation}
Since each $f\in C^\infty(T,A)$ is invariant under integer translations in the $x_1$-direction, $f$ admits a Fourier expansion
\begin{equation}\nonumber
f(x)=\sum_{\gamma^2\in\mathbb Z} g(x^2+\gamma^2) q^{n\gamma^2 x^1}
\end{equation}
where $g$ is rapidly decreasing on $\mathbb R$. If $g$ is Gaussian the corresponding $f$ deserves a special notation
\begin{equation}\label{theta}
\theta_t(nx^1,x^2):=\sum_{\gamma^2\in \mathbb Z} e^{-t(x^2+\gamma^2)^2} q^{n\gamma^2 x^1}\, .
\end{equation}
where $t>0$. Notice that (up to an overall factor) $\theta_t$ is indeed a holomorphic theta-function for a suitable complex structure on $T$. This is an instance of complex polarization familiar in the context of geometric quantization where ground states of the system are labeled by holomorphic section of a suitable line bundle. In this picture, the {\it Bohr-Sommerfeld} (BS) polarization also arises by taking the $t\to 0$ limit of $\theta_t(nx^1,x^2)$ which becomes the periodic delta-function supported on lines (the BS leaves) parallel to the $x^2$-axis and corresponding to those values of $x^1$ for which $nx^1dx^2$ is integral. It is then immediate to conclude that the system has $n$ ground states, one for each leaf. In other words, we found an isomorphism
\begin{equation}\nonumber
\mathcal H \cong C^\infty({\rm BS}(A))\,.
\end{equation}
where we denote by ${\rm BS}(A)$ the Bohr-Sommerfeld locus, i.e. the union of all the BS leaves. 
We emphasize that the dimension of the space of ground states ( i.e.\ the number of Bohr-Sommerfeld leaves), depends only on the curvature $dA$ and not on $A$ itself. While the geometric description of the space of states (the location of the BS leaves does depend on the particular choice of $A$, any two such choices are intertwined by the group $SL_2(\mathbb Z)$ of autormorphism of $\mathbb Z^2$ preserving the two-form $dA$.

\subsection{Twisted  tori: $dB=0$}\label{flat}
We would like to repeat a similar analysis for the sigma model on a three dimensional torus $\tilde X:=\mathbb Z^3\backslash \mathbb R^3$. This means that instead of quantizing a particle constrained on a charged torus, we seek to quantize the free loop space $\mathcal L \tilde X$ in presence of a magnetic field. A $U(1)$-gauge field on $\mathcal L \tilde X$ descends to a B-field $B\in \Omega^2(\tilde X)$. In mathematical jargon one says that such a gauge field defines the structure of a $U(1)$-gerbe on $\tilde X$. The case $dB=0$ has been treated exhaustively in \cite{kapustinorlov}. The main obstacle is represented by the fact that $\mathcal L \tilde X$ breaks up into disconnected components $\mathcal L_w\tilde X$, or winding-sectors, labeled by the set $\pi_1(\tilde X)^\sim$ of conjugacy classes of the fundamental group of $\tilde X$. The latter happens to be commutative and thus coincides with the set of its conjugacy classes. At the level of zero-modes, which is our main concern here, the phase space $\mathcal M$ of our theory decomposes as
\begin{equation}\nonumber
\mathcal M = \coprod_{w\in \mathbb Z^3} \mathcal M_w
\end{equation}
where $\mathcal M_w\cong T^*\tilde X$ for all $w$. The canonical quantization of $\mathcal M$, requires choosing a natural Hilbert space of states $\mathcal H_{\tilde X,w}^0$ for each $w\in \mathbb Z^3$.  According to \cite{kapustinorlov}, $\mathcal H_{\tilde X,w}^0$ can be identified with a completion of
\begin{equation}\nonumber
C_w^\infty(\tilde X,B):=\{f_w\in C^\infty(\mathbb R^3)\,|\, f_w(x+\gamma)=f_w(x) {\rm Hol}_w(\gamma,x)\, |\, \forall \gamma\in \mathbb Z^3\}
\end{equation}
with holonomy factor 
\begin{equation}\nonumber
{\rm Hol}_w(\gamma,x):=q^{\int_x^{x+\gamma} \iota_w B }
\end{equation}
where integration is performed along the straight-line path in $\mathbb R^3$ that connects $x$ to $x+\gamma$.
Geometrically, $C_w^\infty(M,B)$ can be thought of as the space of sections of a line bundle $\mathcal L_w$ supported on the zero-section of $\mathcal M_w$. The collection of these line bundles should be thought of as the low-energy limit of a single flat line bundle on $\mathcal L\tilde X$. Since ${\rm Hol}_w(\gamma,x)$ is multiplicative, 
\begin{equation}\label{definitionko}
C^\infty(\tilde X,B):=\bigoplus_{w\in\mathbb Z^3} C^\infty_w(\tilde X,B)
\end{equation}
inherits the structure of algebra over $C^\infty_0(\tilde X,B)\cong C^\infty(\tilde X)$. Moreover, imposing
\begin{equation}\nonumber
\lim_{B\to 0} C^\infty (\tilde X,B) = C^\infty (\tilde X, 0)
\end{equation}
one obtains the correct quantization of functions $f_w\in C^\infty(\tilde X,B)$ as multiplication operators. A natural basis of $C_w^\infty(\tilde X,B)$ is given by shifted exponentials
\begin{equation}\nonumber
q^{x^im_i-B_{ij}w^j}
\end{equation}
labeled by classical momenta $(m_1,m_2,m_3)\in (\mathbb Z^3)^*$. Momenta are quantized in the usual way as differentiation operators, independently on the winding mode.

\subsection{Twisted tori: $dB\neq 0$}\label{nonflat}
How to extend the results of \cite{kapustinorlov} if $dB\neq 0$? Let us fix the B-field to be
\begin{equation}\nonumber
B:=x^1dx^2dx^3\,.
\end{equation}
Naively, one might hope that definition \eqref{definitionko} still works in this more general setting. Unfortunately, this is not correct for a number of reasons. One way to see this, is to picture the ground state of a string with winding mode $w\in \mathbb Z^3$ as a straight line in the universal cover $\mathbb R^3$ parallel to $w$. In this limit, the system effectively reduces to a charged particle moving in a plane transversal to the string and the particular choice of such a plane should not matter. However if $w=(1,0,0)$, say, then $\iota_w B$ vanishes on the $x^1=0$ planes but not on any other plane transversal to $w$. In order to get the correct answer, one needs modify the definition of holonomy. by adding a term
\begin{equation}\label{holonomy}
{\rm Hol}_w(\gamma,x): =q^{\int_x^{x+\gamma} (\iota_w B+A_w )}
\end{equation}
where $A_w$ is choosen so that $dA_w=B(w)$. It follows that, schematically,
\begin{equation}\nonumber
d(\iota_wB+A_w)= \det (w \,|\, x\,|\, dx) =\iota _w dB
\end{equation}
which is invariant under shifts in the direction of $w$ of a given transversal plane. Once again, elements of $C_w(\tilde X,B)$ can be thought of sections of a line bundle $\mathcal L_w$. The main difference with the $dB=0$ case is that here $\mathcal L_w$ has nontrivial curvature. Thus we learn that for each non-zero $w\in \mathbb Z^3$, $\mathcal H_{\tilde X,w}^0$ can be identified with the space of states obtained from the the quantization of a magnetic torus. $SL_2(\mathbb Z)$ acts naturally on each magnetic torus, adding exact terms to each $A_w$. In fact, these individual $SL_2(\mathbb Z)$ actions combine into the larger $SL_3(\mathbb Z)$-action on $\mathbb Z^3$ fixing $dB$. In other words, $SL_3(\mathbb Z)$ acts by adding exact terms to both $B$ and $A_w$. 

We now turn our attention to the problem of counting the ground states of the theory. We work In analogy with section \ref{magnetic}, our goal is to quantize simultaneously the magnetic tori associated with each winding sector described above. To this end, we define the BS locus ${\rm BS}(B,A_w)$ as the set of points $x\in \tilde X$ such that $\iota_w B+A_w$ is integral. For concreteness, let us fix $A_w:=w^1x^2dx^3$. Then ${\rm BS}(B,A_w)$ can be described as follows. It is the set of points with coordinates $(x^1,x^2,x^3)$ such that $w^1 x^2\in \mathbb Z$ and $(|w^2|,|w^3|)\in \mathbb Z$ where here $(\cdot,\cdot)$ denotes the greatest common divisor of two non-negative integers. The construction is perhaps best understood as performed in two stages. Since for each $w$ all functions in $C^\infty_w(\tilde X)$ are periodic in the $x^1$ direction we first look for those planes transverse to the $x^1$-axis such that the pull back of $\iota_w B$ to them is integral. We denote by ${\rm BS}(B)$ the union of all these planes. Each plane $V\in {\rm BS}$ now descends to a magnetic subtorus of $\tilde X$ with $U(1)$ connection given by $A_w$. The final result is given by taking the union of the BS loci ${\rm BS}(A_w)$ corresponding to each plane $V\in {\rm BS}$. To summarize, the Hilbert space of zero-modes for the twisted torus is given by
\begin{equation}
\mathcal H_{\tilde X}^0\cong \bigoplus_{w\in \mathbb Z^3} C^\infty({\rm BS}(B,A_w)) 
\end{equation}
As in the $dB=0$ case, ${\rm Hol}_w(\gamma,x)$ is multiplicative and therefore $C^\infty(\tilde X,B)$ is an algebra over $C^\infty (\tilde X)$. This fact can be used to quantize elements of $C_w^\infty(\tilde X, B)$ as multiplication operators acting on $C_w^\infty(\tilde X,B)$. On the other hand, ${\rm Hol}_w(\gamma,x)$ has now nontrivial dependence on $x$ and the quantization of momenta requires modification. The correct definition is 
\begin{equation}\nonumber
P_i(f_w):=(\partial_i f+(\varepsilon_{ijk}^{\rm pol} W^j x^k)f)_w \qquad \forall f_w\in C^\infty_w(\tilde X, B)
\end{equation}
where
\begin{equation}\nonumber
\varepsilon^{\rm pol}(w,x,\gamma) = w^1x^2\gamma^3-w^2x^1\gamma^3+w^3x^1\gamma^2
\end{equation}
for $w,\gamma\in \mathbb Z^3$ and $x\in \tilde X$. We emphasize that the Lie algebra generated by the $P_i$ and the $W_j$ is isomorphic to $\fg$ defined in section \ref{sec:introduction}.

\subsection{The double twisted torus}
Consider the ``polarized'' 2-step nilpotent group $G^{\rm pol}$ with underlying vector space $\mathbb R^3\oplus(\mathbb R^3)^*$ and group law
\begin{equation}\nonumber
(x,x^*)(z,z^*)=(x+z,x^*+z^*+\varepsilon^{\rm pol}(x,z,-))\,.
\end{equation}
$G^{\rm pol}$ is isomorphic to the group $G$ introduced in section \ref{sec:introduction} and the corresponding nilmanifold
\begin{equation}\nonumber
Y^{\rm pol}:=G^{\rm pol}(\mathbb Z)\backslash G^{\rm pol}(\mathbb R)
\end{equation}
is diffeomorphic to the double twisted torus $Y$ of section \ref{sec:introduction}. Smooth functions on $Y^{\rm pol}$ are given by
\begin{equation}\nonumber
C^\infty(Y^{\rm pol})=\left(C^\infty(G^{\rm pol}(\mathbb R))\right)^{G^{\rm pol}(\mathbb Z)} \,.
\end{equation}
Since $Y^{\rm pol}$ is fibered by three-dimensional tori, every function $f\in C^\infty(Y^{\rm pol})$ admits a Fourier decomposition along the fiber direction. 
\begin{equation}\nonumber
f(x,x^*)=\sum_{w\in \mathbb Z^3} f_w(x) q^{ w x^*}\,.
\end{equation}
As a consequence, one obtains an isomorphism of algebras
\begin{equation}\nonumber
C^\infty (Y^{\rm pol})\cong \bigoplus_{w\in \mathbb Z^3} C_w^\infty(\tilde X, B) = C^\infty(\tilde X,B)\,.
\end{equation}
This explains why the full space of ground states for the sigma model with target a torus $\tilde X$ twisted by $B$ can be naturally identified with the space of smooth functions on the double twisted torus $Y^{\rm pol}\cong Y$. To explain the relation with the Bohr-Sommerfeld locus, Notice that by construction, each function in $C^\infty(Y^{\rm pol})$ is periodic in the $x^1$ direction.
Therefore, we are allowed to Fourier expand each $f_w\in C_w^\infty(\tilde X,B)$ as 
\begin{equation}\nonumber
f_w(x)=\sum_{k\in \mathbb Z} f_{w,k}(x^2,x^3) q^{ k x^1}\,. 
\end{equation}
$G^{\rm pol}(\mathbb Z)$-invariance forces
\begin{equation}\nonumber
f_w(x)=\sum_{k\in \mathbb Z/(|w_2|,|w_3|)\mathbb Z}\sum_{\gamma^2,\gamma^3\in \mathbb Z} f_{w,k}(x^2+\gamma^2,x^3+\gamma^3) q^{(k+w^3\gamma^2-w^2\gamma^3) x^1 + w^1\gamma^3 x^2}
\end{equation}
in the limit where $f_{w,k}$ is periodic, $f_w$ becomes a distribution supported on ${\rm BS}(B,A_w)$. 

\subsection{The Heisenberg nilmanifold}
We conclude our analysis by arguing that $C^\infty(Y^{\rm pol})$ admits an alternative description as the space of vacuum vectors for the sigma model with target the Heisenberg nilmanifold $X$ i.e.\  
\begin{equation}\nonumber
\mathcal H_X^0:=\bigoplus_{m\in \pi_1(X)^\sim} \mathcal C_m^\infty (X)
\end{equation}   
with sectors labeled by conjugacy classes of the fundamental group and $C^\infty_m(X)\cong C^\infty(X^)$ for all $m$.  We find it convenient to replace the three-dimensional Heisenberg group $H(\mathbb R)$ by its isomorphic image $H^{\rm pol}$ realized as the group of $3\times 3$ upper triangular matrices with unity along the diagonal. We denote by
\begin{equation}\nonumber
X^{\rm pol}:=H^{\rm pol}(\mathbb Z)\backslash H^{\rm pol}(\mathbb R)\cong X
\end{equation}
the corresponding nilmanifold. In this presentation, an element of $\pi(X^{\rm pol})$ is identified with an integer matrix
\begin{equation}\nonumber
\left(
\begin{matrix}
1 & w^3 & p_1\\
0 & 1 & w^2\\
0 & 0 & 1
\end{matrix}
\right)
\in H^{\rm pol}(\mathbb Z)
\end{equation}\,. 
From this we deduce a canonical identification
\begin{equation}\nonumber
\pi_1(X^{\rm pol})^\sim\cong \bigoplus_{w^2,w^3\in\mathbb Z} \mathbb Z/\mathbb (|w^2|,|w^3|)\mathbb Z
\end{equation}
Furthermore, notice that $X^{\rm pol}$ can be thought of a a circle fibration over a two-dimensional torus, fiberwise Fourier expansion leads to a further decomposition into subsectors
\begin{equation}\nonumber
C_m^\infty(X^{\rm pol}) \cong \bigoplus_{k\in \mathbb Z} C_{m,k}^\infty(X^{\rm pol})\,.
\end{equation} 
If $x_1^*$ is the fiber coordinate and $x^2, x^3$ are coordinates on the base $T_{2,3}$ of the circle fibration of $X^{\rm pol}$, then using the notation of \ref{magnetic}
\begin{equation}\nonumber
C_{m,k}^\infty(X^{\rm pol})\cong C^\infty (T_{2,3},k x^2dx^3)\,. 
\end{equation}
For $k=0$, this is are nothing but all smooth functions on the two-torus $T_{2,3}$. For $k\neq 0$, this decomposes into $|k|$ subsectors corresponding to the Landau levels of the magnetic torus $T_{2,3}$ with $U(1)$ gauge field $kx^2dx^3$. Equivalently,
\begin{equation}\nonumber
\mathcal H_{X^{\rm pol}}^0\cong \bigoplus_{w\in \mathbb Z} C^\infty({\rm BS}(w^1x^2dx^3))\otimes \mathbb Z/(|w^2|,|w^3|)\mathbb Z
\end{equation}
We conclude that the HDFT ground states are labeled by
\begin{equation}\nonumber
\left(\mathbb Z\langle p_3\rangle\bigoplus_{w^2,w^3\in \mathbb Z} \mathbb Z\langle p_1\rangle/(|w^2|,|w^3|) \bigoplus_{w^1\in \mathbb Z} \mathbb Z\langle p_2\rangle/|w^1|\mathbb Z \right)\, .
\end{equation}
At the level of ground states, we see that $T$-duality exchanges $w^1$ and $p_1$ i.e.\ winding and momentum in the direction dualized, as expected. We point out that the cooridinates being dualized are $x^1$ and $x_i^*$ as a consequence of our choice of B-field $x^1dx^2dx^3$. Choosing $-x^2dx^1dx^3$ would lead to dualizing $x^2$ and $x_2^*$, which makes the connection with the calculations of section \ref{sec:5}.

\end{document}